\documentclass[prb,aps,amsmath,amssymb,superscriptaddress,longbibliography,notitlepage,twocolumn]{revtex4-1}
\usepackage{graphicx,multirow,outlines,ulem}
\usepackage{color}
\usepackage{hhline}
\usepackage{physics}
\usepackage{hyperref}

\usepackage[mathscr]{eucal}

\usepackage{tikz}
\usepackage{xcolor}

\newcommand{\bo}{\begin{outline}}
\newcommand{\eo}{\end{outline}}

\newcommand{\qed}{\nobreak \ifvmode \relax \else
      \ifdim\lastskip<1.5em \hskip-\lastskip
      \hskip1.5em plus0em minus0.5em \fi \nobreak
      \vrule height0.75em width0.5em depth0.25em\fi}

\begin{document}

\title{Measurement-induced phase transition in periodically driven free-fermionic systems}
\author{Pallabi Chatterjee}
\email{ph22d001@iittp.ac.in }    
\author{Ranjan Modak}
\email{ranjan@iittp.ac.in}    
\affiliation{Department of Physics, Indian Institute of Technology Tirupati, Tirupati 517619, India} 
\begin{abstract}
It is well known that unitary evolution tends to increase entanglement, whereas continuous monitoring counteracts this growth by pinning the wavefunction trajectories to the eigenstates of the measurement operators. In this work, we investigate the fate of the measurement-induced phase transition in a periodically driven free-fermionic quantum system, where the hopping amplitude is modulated periodically in time using a square pulse. 
In the high-frequency limit, a renormalization group analysis of the non-Hermitian quantum sine-Gordon model  [as proposed in \href{https://doi.org/10.1103/PhysRevX.11.041004}{Phys. Rev. X 11, 041004 (2021)}]  reveals \textcolor{black} {that if the hopping amplitude is varied symmetrically around zero, the system always favors the area-law phase, where the steady-state entanglement entropy is independent of subsystem size. In contrast, asymmetry in the drive amplitudes tends to promote entanglement growth. Furthermore, numerical evidence for the system sizes accessible to us suggests that decreasing the drive frequency typically favors entanglement growth. For such driven systems, at least for reasonably small frequency regimes, as a function of measurement strength, we observe a potential signature of a Berezinskii-Kosterlitz-Thouless (BKT) phase transition between a gapless critical phase, characterized by logarithmic growth of entanglement entropy with subsystem size, and a gapped area-law phase. However, it is almost impossible to rule out the possibility that the transition observed here is not an actual thermodynamic transition, but a finite-size crossover between logarithmic to area law entanglement phase. 
Even in that scenario, the critical length scale beyond which the area law phase prevails increases with the increasing time period of driving. On the other hand, for a symmetric drive, the system consistently exhibits an area-law phase, regardless of the driving frequency.}



\end{abstract}
\maketitle

\section{Introduction}

Understanding the entanglement properties of quantum many-body systems has become a key focus of research, particularly in relation to different phases of matter and the transitions between them, as well as identifying quantum chaos at the many-body level~\cite{vidmar_marcos.17,vidmar.17,vidmar.prl.18,lucas.19,srednicki.19,leblond.19,nag.2020,eisert2010colloquium,modak2021eigenstate}. Thanks to recent progress on the direct measurement of entanglement dynamics with cold atoms and trapped ions, a key question of how entanglement spreads in out-of-equilibrium many-body systems brought lots of attention in the last decade~\cite{islam2015measuring,kaufman2016quantum}.
Entanglement entropy, a quantifier of quantum correlations between subsystems, exhibits distinct scaling behaviors that provide insight into the nature of the underlying phase.
For example, in gapped quantum system, the ground state entanglement entropy typically obeys an area law (scales with the boundary area of the subsystem); in contrast, in one-dimensional critical systems, where long-range correlations are prevalent, it follows a logarithmic scaling with subsystem size, reflecting the scale-invariant nature of these systems at criticality~\cite{osterloh2002scaling,calabrese2004entanglement,pollmann2009theory}.
On the other hand, under unitary evolution, usually for clean short-ranged systems, the entanglement growth is algebraic in time before saturating to a value that scales linearly with the subsystem size, following a volume law~\cite{alba2017entanglement,calabrese2004entanglement,chapman2019complexity,modak2020entanglement}. 
However, introducing disorder into systems changes the entanglement behavior drastically. In a closed, non-interacting quantum system, even small amounts of disorder—whether in one dimension or two—can prevent the thermalization of single-particle states, leading to a phenomenon known as Anderson localization~\cite{PhysRev.109.1492}. Furthermore, when interactions are introduced into these disordered systems, the many-body localized (MBL) phase emerges~\cite{BASKO20061126,PhysRevB.75.155111,PhysRevB.77.064426,PhysRevB.82.174411,PhysRevX.5.041047,PhysRevLett.115.230401,PhysRevResearch.3.033043,PhysRevLett.109.017202}. In the MBL phase, entanglement grows logarithmically over time, despite the absence of energy transport~\cite{moore.2012,dima.2013,garg.22}, in stark contrast to thermal systems, where entanglement growth is typically linear~\cite{kim.2013}. Moreover, the typical eigenstate entanglement entropy behaves quite differently in the thermal (ergodic) phase, which follows volume law scaling, compared to the non-ergodic MBL phase, where the scaling follows an area law~\cite{alet.2015}. Studies of the fate of MBL are not limited to closed systems; there are several interesting results reported where the system is coupled to a bath~\cite{PhysRevLett.123.030602,PhysRevX.7.011034,PhysRevLett.116.237203,PhysRevLett.116.160401,PhysRevB.93.094205,PhysRevLett.123.090603,PhysRevLett.124.130602}.

Another class of entanglement phase transitions arises from the continuous monitoring of quantum systems. The stochastic Schrödinger equation typically governs these open quantum dynamics, and averaging over different realizations of the states leads to the well-known Lindblad master equation dynamics~\cite{paule2018thermodynamics}. Since entanglement entropy is a nonlinear function of the state, calculating it for each trajectory separately and then averaging reveals a phase transition. In contrast, any linear function of the state is featureless and does not exhibit such transitions. Several studies have explored these entanglement phase transitions in both interacting and non-interacting systems, considering clean as well as disordered systems with long-range and short-range hopping, using both Lindblad jump operators and quantum state diffusion approaches~\cite{diehl_prl_2021, areejit_prb,cao2019entanglement,PhysRevLett.128.010603,PhysRevLett.128.010605,PhysRevB.107.L220201,PhysRevB.102.054302,Ladewig_2015}. \textcolor{black}{Several studies have also been reported on the Ising chain with $Z_2$  symmetry under weak measurements accompanied by quantum jumps, demonstrating a transition between a critical phase and a Zeno phase~\cite{Ising_1, Ising_2, Ising_3}. 
The impact of a single quantum jump on a localized site in a free fermionic system with $U(1)$ symmetry has been explored in Ref.~\cite{di2024entanglement}, along with investigations into two-particle drive and loss processes in Ref.~\cite{single_site_2}. 
Additionally, the effect of different unravelings of the Lindblad master equation—namely, quantum state diffusion and quantum jumps—on the Ising chain is discussed in Ref.~\cite{diff_unravelling_Ising}. Interestingly, these two distinct measurement schemes lead to different patterns of entanglement growth. Further studies have examined the behavior of entanglement under long-range measured operators~\cite{long_range_operator_1, long_range_operator_2}, as well as the delocalization properties and bifurcation characteristics in the distribution of the expectations of the measured operators for a free fermionic system with $U(1)$ symmetry~\cite{bifurcation_piccitto2024impact}.} Continuously monitored open systems display competition between unitary evolution, which drives entanglement growth, and measurements, which tend to localize the wavefunction by collapsing it into eigenstates of the measurement operators. As a result, these systems undergo a measurement-induced phase transition as a function of measurement strength, which also has been reported recently in experiments~\cite{koh2023measurement}. It has been shown that, in both clean and disordered free-fermionic systems, there is an entanglement phase transition from a critical to an area law phase as the strength of continuous measurements is tuned. This transition was predicted to fall into the Berezinskii-Kosterlitz-Thouless universality class~\cite{cao2019entanglement,areejit_prb,diehl_prl_2021}, a result that was also supported by a theoretical study in which the dynamics of a continuously monitored free fermionic system were described by an effective non-Hermitian sine-Gordon model~\cite{PhysRevX.11.041004}. 
However, it is important to point out that the stability of the critical phase in a one-dimensional free-fermionic system has been questioned recently, and it has been argued in the thermodynamic limit, for any finite measurement strength, only the area-law phase will survive~\cite{mirlin.2023,starchl2024generalized, Interacting_NLSM_1} for random projective and quantum jump measurement protocols. \textcolor{black}{Also in the case of Dirac fermions, the same conclusion is claimed recently\cite {arealaw_2_muller2025monitored} after refining their previous results\cite{PhysRevX.11.041004}. There are recent theoretical studies of a monitored chain of a system of size $L$ of Majorana fermions, which shows ($\ln L)^2$ behavior in the critical phase~\cite{fava2023nonlinear}, in contracts for noninteracting systems where the $U(1)$ symmetry is preserved, reported to show area-law phase for all measurement strength in presence of complex random hopping\cite{complex_hopping}.} 
On the other hand, in the case of interacting systems, a transition from a volume law to an area law has been observed~\cite{PhysRevB.102.054302,PhysRevB.107.L220201,fazio.prbl.24}. Furthermore, the role of measurements in localizing quantum states in nonunitary quantum circuits has also attracted significant interest~\cite{PhysRevLett.128.050602,PhysRevLett.128.130605,PhysRevB.107.014308,PhysRevLett.131.220404,PhysRevLett.128.240601,PhysRevLett.128.010604,PhysRevResearch.3.023200,PhysRevB.104.155111,PhysRevB.104.104305,PhysRevB.103.174309,PhysRevB.101.060301,PhysRevB.102.014315,PhysRevLett.125.070606,PhysRevX.10.041020,PhysRevB.101.104301,PhysRevLett.125.030505,PhysRevResearch.2.013022,PhysRevB.101.104302}. Some of these studies have also been extended in dimension $D>1$ ~\cite{Chahine_2024_2D,singleparticle_measurement_2024,Poboiko_2d_PhysRevLett.132.110403} and a semi-classical limit of such transition has also been introduced recently~\cite{sumilan_2024_measurement}. \textcolor{black}{In the case of $D > 1$, Ref.~\cite{Chahine_2024_2D} maps a two-dimensional free fermionic monitored lattice system to a nonlinear sigma model (NLSM), suggesting that a $D$-dimensional free fermionic measurement problem can be mapped to a $(D+1)$-dimensional Anderson localization problem. If this mapping holds in one dimension, it implies that there would be no measurement-induced phase transition in the quantum state diffusion (QSD) protocol used in this manuscript. Additionally, for the two-dimensional bipartite lattice, the authors also map the monitored system to a non-Hermitian Hubbard model and observe numerical results that deviate from the predictions of the NLSM-based Anderson localization model. Specifically, they report the emergence of a conformal field theory (CFT) at the critical point—behavior that is absent in the Anderson localization scenario described by the NLSM. Since the Hubbard model mapping is valid for the continuous QSD protocol, the study raises important questions about the correspondence between the projective and QSD protocols, and about the broader validity of mapping a $D$-dimensional measurement problem to a $(D+1)$-dimensional Anderson localization problem. Notably, the projective measurement protocol shows strong parallels with the $(D+1)$-dimensional Anderson model applied to a $D$-dimensional free fermionic lattice system~\cite{2D_projective}. }
In contrast to disorder-induced localization observed in the Anderson-localized or MBL phase, the measurement-induced localization is governed by a completely different mechanism and driven by quantum measurements and measurement strength.

\begin{figure}
\centering
\includegraphics[width=0.48\textwidth]{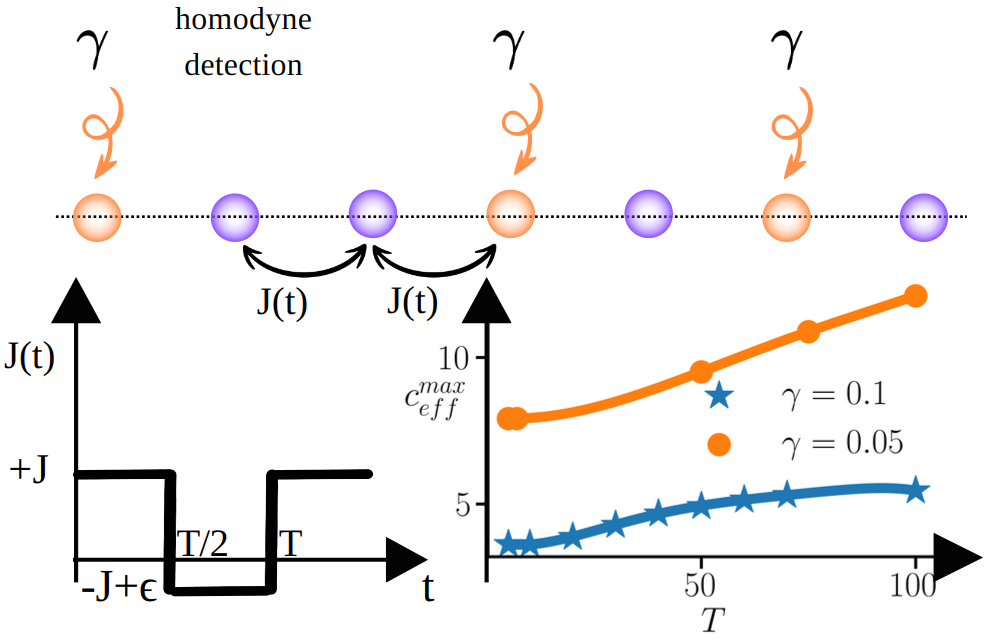}
    \caption{Schematic of our protocol, where a free-fermionic lattice is continuously monitored, and the hopping amplitude $J(t)$ is periodically varied. Data is for the maximum of the local effective central charge vs time-period $T$ for different measurement strength $\gamma$ with $\epsilon=3J/4$. }
    \label{fig0}
\end{figure}

In parallel, in the last few decades, periodically driven (Floquet) quantum systems have emerged as a powerful framework for studying non-equilibrium phases of matter~\cite{PhysRevB.109.094306,PhysRevB.103.184309,PhysRevB.105.024301,PhysRevB.89.165425,PhysRevB.93.174301,PhysRevB.97.020303,Aditi_2024}. Floquet systems are known to exhibit exotic phases, such as discrete-time crystals~\cite{PhysRevLett.116.250401,PhysRevLett.117.090402,zhang2017observation} and Floquet topological insulators~\cite{PhysRevB.79.081406,lindner2011floquet,PhysRevB.84.235108,PhysRevLett.113.236803,PhysRevB.94.155122}. The fate of many-body localization (MBL) in the presence of a periodic drive has also been studied~\cite{PhysRevLett.115.030402,dima.prl}.
 While it has been demonstrated that Floquet systems can stabilize or even induce new phases, it remains an open question how continuous monitoring affects these driven systems and whether the introduction of a periodic drive can significantly alter the dynamics of a continuously monitored system, either enhancing or suppressing the effects of measurements.
The effect of time-dependent Hamiltonian has been studied within the Lindblad master equation formalism~\cite{PhysRevA.99.022105,10.21468/SciPostPhysCore.4.4.033}, but the effect of periodic drive on a continuously monitored open quantum system, particularly in terms of trajectory-averaged entanglement entropy, remains \textcolor{black}{relatively} unexplored\cite{drive_qcircuit_PhysRevB.100.134306}. In our study, we investigate such cases, and our main finding, based on our accessible system size numerics, is that periodically driven non-interacting systems tend to exhibit a signature of undergoing a BKT transition for reasonably high driving periods. In general, for a given drive protocol, the low-frequency regime favors the critical phase more, implying that if the BKT transition persists, then the critical measurement strength (transition point) increases with the time period. 
Moreover, the symmetric drive of the hopping amplitude about a zero mean is special; such a system always favors the
area-law phase in the thermodynamic limit, independent of the frequency regime, in contrast to what has been observed in the case of the disorder-induced scenario~\cite{PhysRevB.103.184309}. We summarize our setup and main results based on the finite-sized numerical data in the form of a schematic in Fig.~\ref{fig0}, which includes real data. \textcolor{black}{We plot the maximum of the local effective central charge, which is a measure of the maximum growth rate of the steady state entanglement with sub-system size (defined later in the manuscript in Eq.~\eqref{c_eff_eq}) with time-period $T$ for different measurement strength $\gamma$. 
\textcolor{black}{It is increasingly believed, based on recent analytical studies, that in the absence of drive, even an infinitesimal measurement strength is sufficient to stabilize the area-law entanglement phase in the thermodynamic limit. In particular, analyses using the nonlinear sigma model (NLSM) have shown that the measurement-induced phase transition in one-dimensional free fermionic lattice models is bounded by a characteristic length scale, which grows exponentially as $\gamma$ decreases. However, clear numerical evidence remains lacking—especially for the specific measurement protocols considered in this work—due to the exponentially diverging correlation length. That said, for quantum jump and projective measurement protocols, numerical signatures consistent with area-law entanglement at small $\gamma$ have been reported~\cite{arealaw_1_starchl2024generalized,mirlin.2023}.} Moreover, to the best of our knowledge, no analytical prediction is available for driven systems till now. This work aims to do a detailed comparative numerical analysis between the driven system with a reasonably low-frequency regime (in the main text) and the no-drive/high-frequency regime (in the Appendix~\ref {nodrive_central_charge}) results for a finite-sized system.}

This manuscript is structured as follows. In Sec .~\ref {model}, we describe our system and protocols. Then, to build the intuitions about our systems, we focus on a $2\times2$ toy model in Sec .~\ref {toy}. Section~\ref {manybody} is dedicated to the many-body systems, which consist of the analytical prediction for the high-frequency limit in sec.~\ref{analytic}, as well as the detailed numerical results of the dependency of entanglement growth on driving periods in sec.~\ref{numerics}. Section~\ref {e0} concentrates on the fate of entanglement entropy growth on changing the asymmetry parameter $\epsilon$ of the square pulse drive as well as the fate of the sinusoidal driving protocol. Finally, we summarize our results in sec.~\ref{conclusions}.
\section{Model and Methodology \label{model}}
We consider spinless fermions in a periodic tight-binding lattice of the length $L$ where the hopping amplitude changes periodically with time, and the system is continuously monitored. The Hamiltonian of this system is given by:
\begin{equation}
    H(t)=J(t)\sum_{i=1}^L \hat{c}_i^{\dag}\hat{c}_{i+1} + h.c.,
    \label{hamiltonian}
\end{equation}
where, $\hat{c}^{\dag}_i$ and $\hat{c}_{i}$ are real-space fermionic creation and annihilation operators respectively. We study the effect of two different kinds of driving with the time period (frequency) $T$ ($\omega=1/T$): one is square pulse drive, where $J(t)= +J$ $(-J+\epsilon)$ for $0\le t < T/2$ ($T/2\le t < T$), with $\epsilon \in$
$[0, 2J]$ is an asymmetry parameter in the drive amplitude (see Fig.~\ref{fig0}); the other is the sinusoidal drive where $J(t)=J\sin(2\pi t/T)$. Note that in the limit $\epsilon=0$, the mean of $J(t)$ is $0$ over a time period, and identify it as a symmetric drive. We examine this limit separately in this manuscript. 
We primarily present the results for the square pulse in the subsequent sections of the manuscript; only at the end, in Sec.~\ref{e0}, some results for the sinusoidal drive are presented. For all our numerical calculations, we choose $J=1/2$. The time evolution of the state can be described using the stochastic Schrodinger equation as
\begin{eqnarray}
    d|\psi(t)\rangle=-iH(t) dt |\psi(t)\rangle&-\frac{\gamma dt}{2}\sum_i (\hat{n}_i-\langle \hat{n}_i \rangle)^2|\psi(t)\rangle\nonumber\\
    &+\sum_i(\hat{n}_i-\langle \hat{n}_i \rangle)dW_i^t |\psi(t)\rangle.\nonumber\\
\label{nlse}
\end{eqnarray}

The above equation describes the continuous monitoring of the particle number $\hat{n}_i=\hat{c}_i^{\dag}\hat{c}_i$ on each site, with measurement strength $\gamma$~\cite{cao2019entanglement}. $dW_i$s are taken from uncorrelated normal distributions having mean $0$ and variance $\gamma dt$, which implies,
$\overline{dW_{i,t}}=0$ ~~~ and ~~~$\overline{dW_{i,t}dW_{j,t'}}=\gamma dt \delta_{i,j}\delta(t-t')$. \textcolor{black}{In our study, we use the quantum state diffusion protocol, while there are similar studies using the quantum jump protocol in clean as well as in disordered systems~\cite{diehl_prl_2021, matsubara2025measurement_quasiperiodic, arealaw_1_starchl2024generalized}.}
For our model and protocols, the total number of particles $N=\sum_i\langle \hat{n}_i\rangle$ remain conserved in the entire time evolution process; hence, if we start with an initial state with a particular filling at $t=0$, for any time $t>0$, the filling fraction remains the same. In our work, we have fixed the filling fraction $N/L=1/2$ for all many-body calculations.  Also, note that the last two terms in Eq.~\eqref{nlse} describe the contribution from the continuous measurement process and can be realized by homodyne detection in quantum optics~\cite{PhysRevA.47.642,oksendal2003stochastic}.

In order to solve Eq.~\eqref{nlse}, one needs to perform Trotterization and can write,
\begin{equation}
    |\psi(t+dt)\rangle=\mathcal{N}e^\mathcal{M}e^{-iH(t)dt}|\psi(t)\rangle,
    \label{trot_wave}
\end{equation}
where, $\mathcal{M}=\sum_i[dW_i^t+(2\langle \hat{n}_i \rangle-1)\gamma dt]\hat{n}_i$, and $\mathcal{N}$ is the normalization constant. Any pure Gaussian state can be represented as~\cite{cao2019entanglement}, 
\begin{equation}
    |\psi\rangle=\prod_{i=1}^N(\sum_{j=1}^L U_{ji}\hat{c}_j^\dag)|0\rangle.
\end{equation}
 As $|\psi(t)\rangle$ evolves according to Eq.~\eqref{trot_wave}, and it preserves the Gaussianity of the state, hence; it is straight forward to show that the $U$ evolves as,
\begin{equation}
    U(t+dt)=e^Me^{-iH(t)dt}U(t),
\end{equation}
where $H(t)=J(t)\sum_{i=1}^L (\hat{c}_i^{\dag}\hat{c}_{i+1} + h.c.)
$ is the single particle Hamiltonian and $M_{ij}=\delta_{i,j}[dW_i^t+(2\langle \hat{n}_i \rangle-1)\gamma dt]$. Under this non-Hermitian evolution, we need to properly normalize the wavefunction in each step, which can be done using the $QR$ decomposition of the $U$ matrix in each step as $U(t+dt)=QR$, and we set the new $U$ as $Q$. \textcolor{black}{Since the columns of $Q$ remain orthonormal, this ensures that the quantum state remains normalized (see Appendix.~\ref{QR decomposition}).} Note that $U$ is an $L\cross N$ matrix, follows $U^{\dag}U=\mathbb{I}_{N\cross N}$. On the other hand, the correlation matrix $D$ can be defined as $D_{ij}=\langle \hat{c}_i^{\dag}\hat{c}_j\rangle=[UU^{\dag}]_{ij}$ . For our many-body calculations, we choose $dW_i$'s from different random realizations, compute the nonlinear function of states, e.g., entanglement entropy, using the correlation matrix for each realization separately, and finally average over different realizations (we ensure the convergence of our results by varying the number of realizations). On the other hand, the evolution of the average state will be governed by the Lindblad equation~\cite{cao2019entanglement}.

\section{$2\times2$ Toy model\label{toy}}
\begin{figure}[!h]
    \centering
    \includegraphics[width=0.48\textwidth]{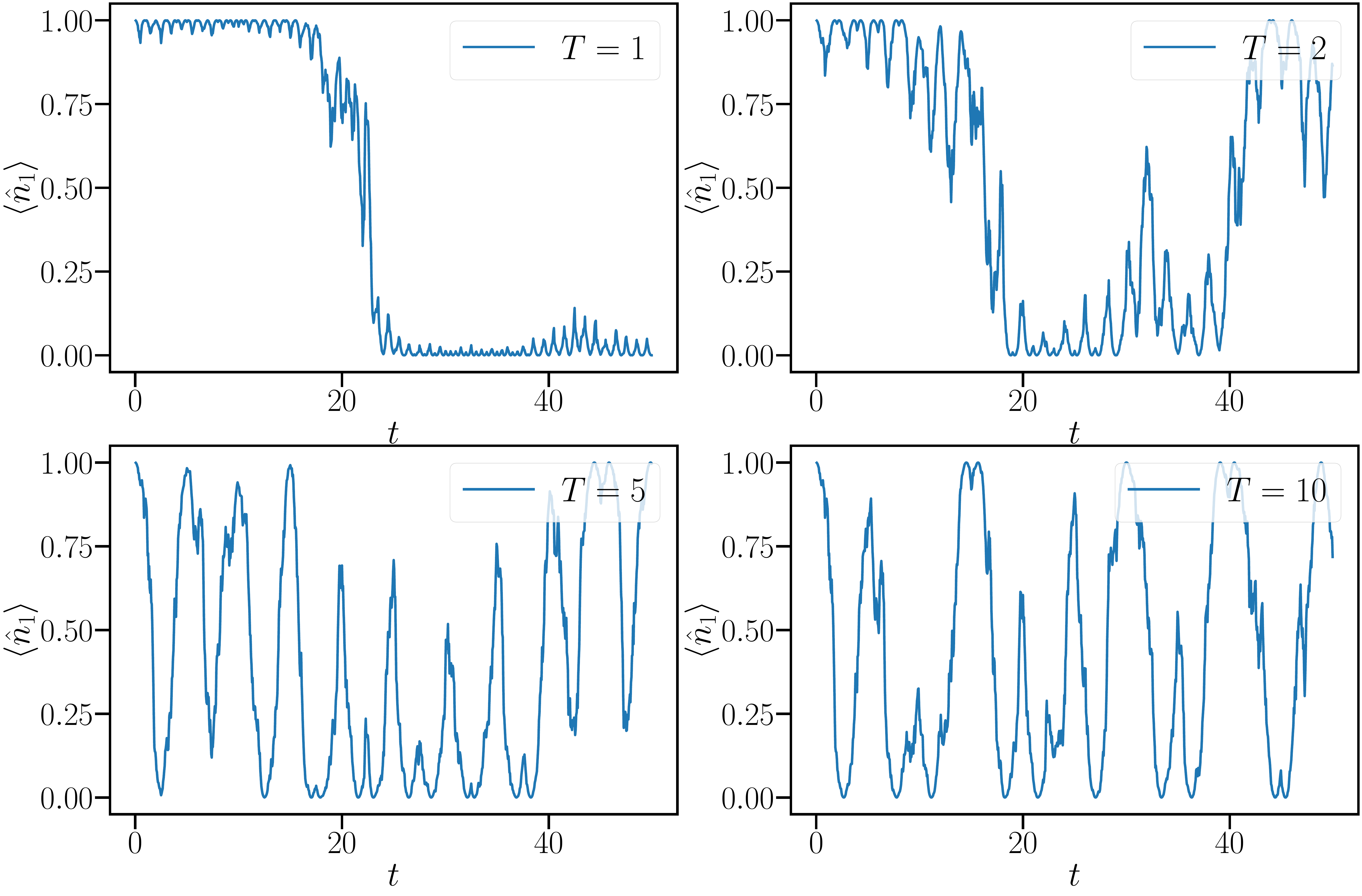}
    \caption{The variation of the expectation value of fermionic occupancy, \( \langle \hat{n}_1 \rangle \), at site 1 as a function of time \( t \) for a typical single trajectory in a \( 2 \times 2 \) periodically driven toy model, for different values of \( T \) and with \( \epsilon = 0 \) fixed.}
    \label{figa}
\end{figure}
\begin{figure}[!h]
    \centering
    \includegraphics[width=0.48\textwidth]{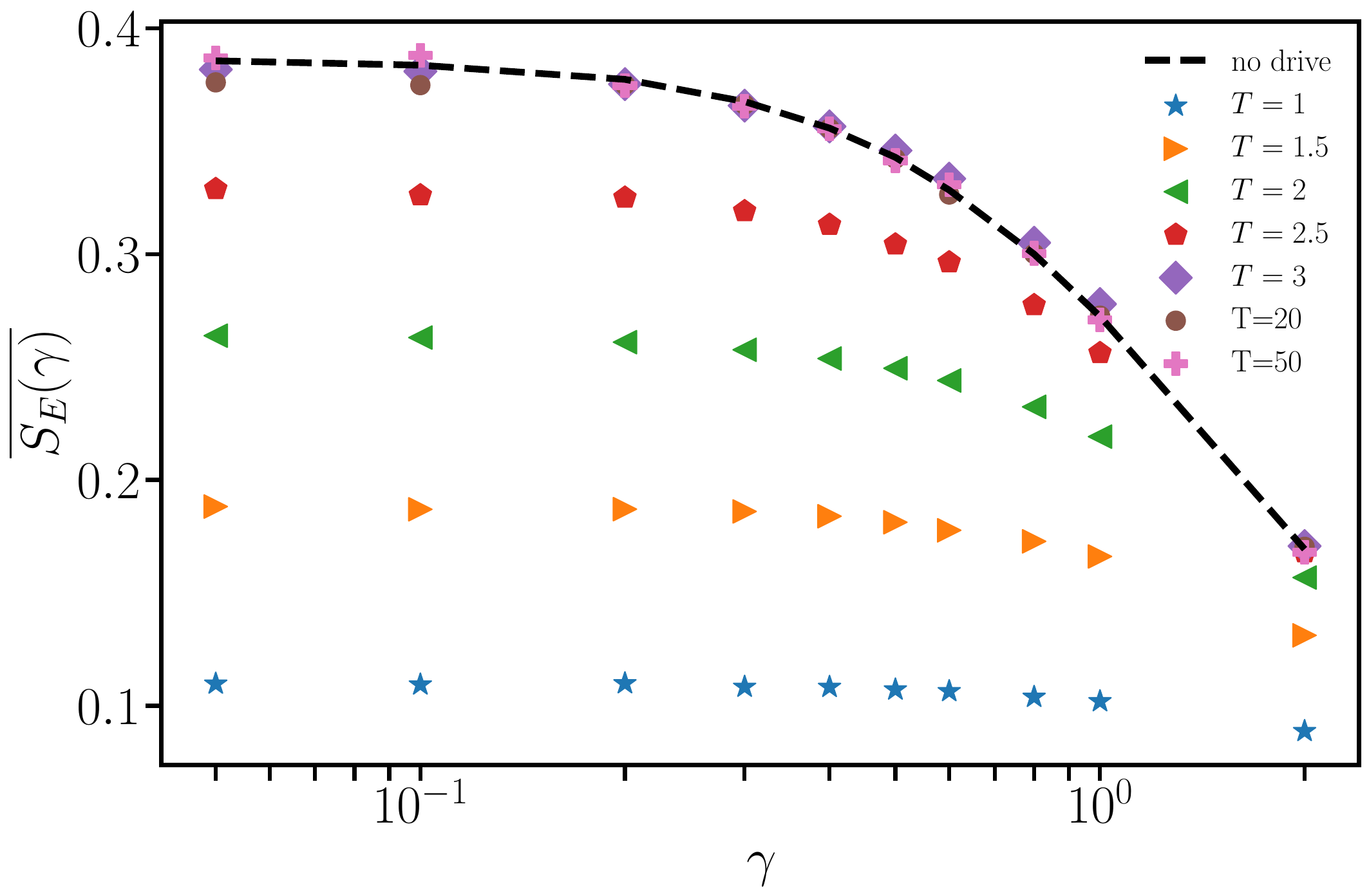}
    \caption{ $\overline{ S_E(\gamma)}$ vs.$ \gamma$ plots for different driving periods for $\epsilon=0$ at steady state. The black dashed line corresponds to no-drive results with $J(t)=J=1/2$. }
    \label{figb}
\end{figure}
\begin{figure}[!h]
    \centering
    \includegraphics[width=0.48\textwidth]{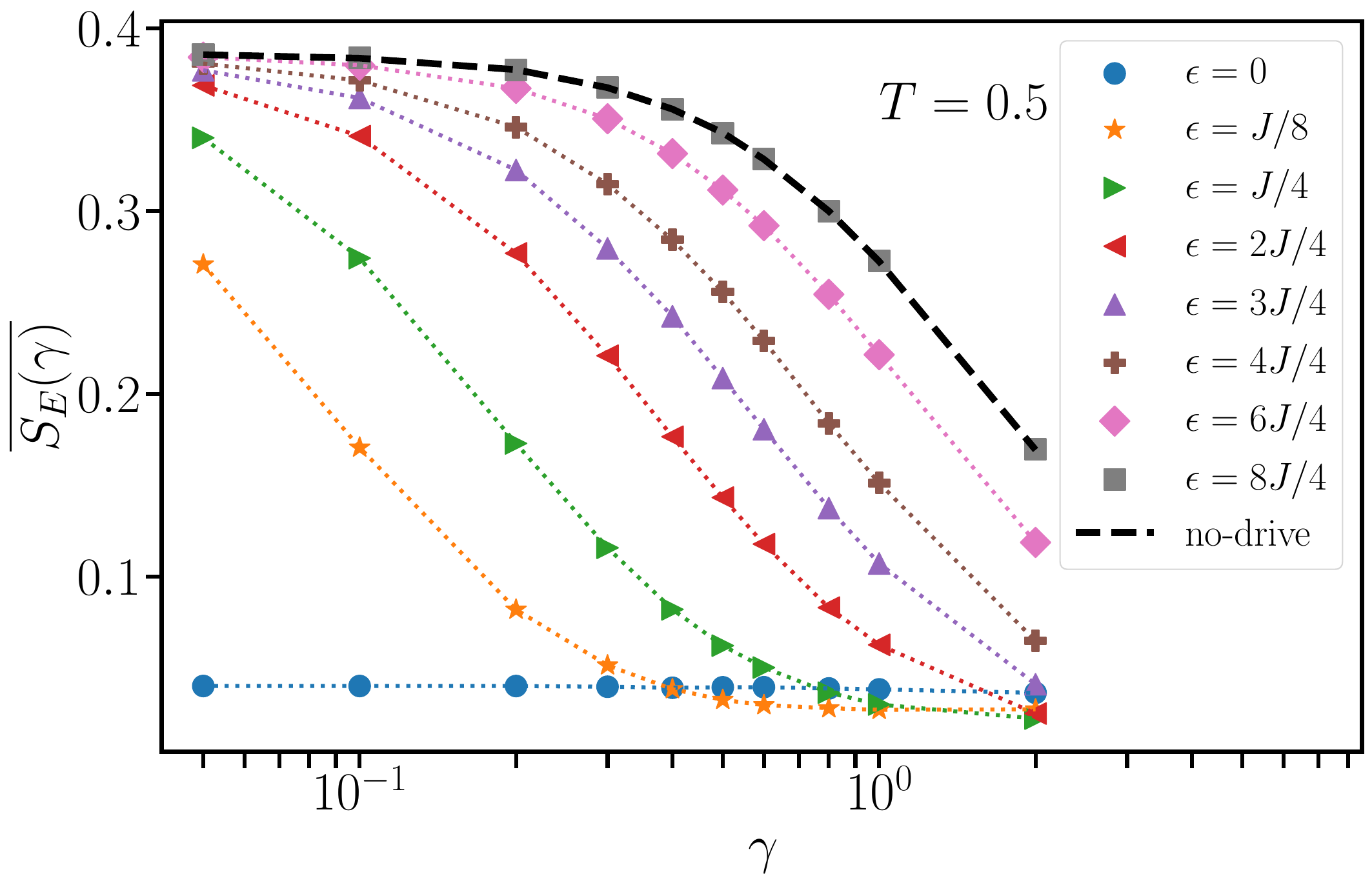}
    \caption{ $\overline{ S_E(\gamma)}$ vs.$ \gamma$ plots for different $\epsilon$ values, for $T=0.5$ at steady state. In the limit $\epsilon=2J$, we get back the no-drive($J(t)=J=1/2$) result, which is represented by the black dashed line. }
    \label{fig_epsilon}
\end{figure}

In this section, we study a periodically driven, continuously monitored $2 \times 2$ system to gain insight into the effect of driving on the system under a square pulse driving protocol. We use the same Hamiltonian as in Eq.~\ref{hamiltonian} and the same protocol as described in Sec.~\ref{model} but limit ourselves to two sites and one particle. In this system, the Hilbert space is spanned by the basis states $|01\rangle$ and $|10\rangle$, which are also the eigenstates of the measurement operators $\hat{n}_1$ and $\hat{n}_2$. In Fig.~\ref{figa}, we plot the expectation value of the occupancy $\langle \hat{n}_1 \rangle$ at site $1$ for a typical single trajectory as a function of time, for different driving periods $T$, but for the fixed $\gamma = 0.1$ and $\epsilon = 0$. For very small $T$, i.e., in the high-frequency regime, $\langle \hat{n}_1 \rangle$ remains mostly at either $0$ or $1$, indicating that the particle is pinned or localized in the eigenstates of the measurement operator or at the dark states, even for a small measurement strength $\gamma = 0.1$. In contrast, as we increase $T$, the particle becomes depinned from the dark states.

To quantify this behavior more systematically, we define a measure,
\begin{eqnarray}
    S_E(\gamma)={ -\langle\hat{n}_1\rangle\ln\langle\hat{n}_1\rangle-\langle\hat{n}_2\rangle\ln\langle\hat{n}_2\rangle }.
\end{eqnarray}
 $S_E(\gamma)$ is averaged over different random trajectories, followed by a long-time average, which we denote as $\overline{S_E(\gamma)}$. In the localized or pinned phase, where both $\langle \hat{n}_1 \rangle$ and $\langle \hat{n}_2 \rangle$ remain at $0$ or $1$ most of the time, the steady-state value of $\overline{S_E(\gamma)}$ tends to zero. In contrast, in the delocalized phase, this value becomes significantly greater than zero. In Fig.~\ref{figb}, we plot the steady-state value of $\overline{S_E(\gamma)}$ as a function of $\gamma$ for different driving periods. For a very small driving period, $T = 1$, $\overline{S_E(\gamma)}$ is almost zero for all $\gamma$. As we increase $T$, the steady-state value $\overline{S_E(\gamma)}$ gradually increases at-least for small $\gamma$. This behavior indicates a potential transition from depinning to pinning as $\gamma$ increases, particularly for large $T$. Interestingly, for $T \gtrsim 2.5$, the steady-state value $\overline{S_E(\gamma)}$ stabilizes and shows almost negligible change upon increasing $T$ further. Moreover, $T \gtrsim 2.5$ data is almost indistinguishable from the no-drive results.
 This gives us a hint that large $T$ promotes depinning, and also, for $\epsilon=0$, the large  $T$ limit may resemble a no-drive scenario even for many-body systems. 

While Fig.~\ref{figb} focuses on the case with $\epsilon = 0$, we now examine the dependence of $\overline{S_E(\gamma)}$ on $\epsilon$ in the high-frequency regime with $T = 0.5$. Figure~\ref{fig_epsilon} shows that for $\epsilon = 0$, $\overline{S_E(\gamma)}$ remains nearly zero for all $\gamma$, as also observed in Fig.~\ref{figb}. As $\epsilon$ increases, we observe a typical increase in $\overline{S_E(\gamma)}$. However, with further increases in $\gamma$, $\overline{S_E(\gamma)}$ approaches zero even for finite $\epsilon$, indicating a transition from depinning to pinning. From Fig.~\ref{fig_epsilon}, it is also evident that the decay of $\overline{S_E(\gamma)}$ to zero occurs more rapidly for smaller $\epsilon$ compared to larger $\epsilon$. This suggests, indirectly, that the critical measurement strength $\gamma_c$ (if a measurement-induced transition exists) may decrease as $\epsilon$ decreases, favoring the localized phase more. In the following sections, we will now focus on the many-body Hamiltonian.

\section{Many-body system\label{manybody}}

In this section, we focus on the periodically driven many-body system described by the Hamiltonian in Eq.~\eqref{hamiltonian}, which is continuously monitored. We initialize the system in a Néel state \( |\psi_0\rangle = |1010\cdots\cdots1010\rangle \) at time \( t = 0 \) and let the system evolve according to Eq.~\eqref{nlse}. A system undergoing a phase transition could be characterized using the scaling of entanglement entropy with system size~\cite{skinner2019measurement,minato2022fate,fava2023nonlinear}. In our work, we also use entanglement entropy as a diagnostic tool.

The entanglement between subsystem \( A \) of length \( l \) and the rest of the system is measured by the von Neumann entanglement entropy, which is defined as \( S = -\text{tr}(\rho_A \ln \rho_A) \)~\cite{fazio.rmp}, where the reduced density matrix \( \rho_A \) is obtained by tracing over the degrees of freedom of the rest of the system (which we identify as \( B \)), i.e., \( \rho_A = \text{tr}_B |\psi(t)\rangle \langle \psi(t)| \). 
Even in the continuously monitored driven system we are interested in, the evolution generators are quadratic, preserving the Gaussianity of the quantum trajectories. Hence, the entanglement entropy \( S \) can be easily calculated using the eigenvalues \( \lambda_i \) of the correlation matrix \( D_{ij} = \langle\hat{c}_i^{\dag} \hat{c}_j \rangle \) as~\cite{peschel2003calculation,peschel2009reduced,vidmar.17},
\begin{equation}
S(l, L) = -\sum_{i=1}^{l} \left[ \lambda_i \ln \lambda_i + (1 - \lambda_i) \ln (1 - \lambda_i) \right].
\label{ent_la}
\end{equation}

\textcolor{black}{In volume law phase, entanglement entropy increases linearly with subsystem size $l$; on the other hand, in area law phase, it is independent of the subsystem size. CFT with periodic boundary predicts the growth of the entanglement entropy in the critical phase as~ \cite {diehl_prl_2021}, 
 \begin{equation}
     S(l,L)=\frac{c(\gamma)}{3}\ln (\frac{L}{\pi}\sin \frac{\pi l}{L}) + S_0(\gamma),
 \end{equation}
 where $ c(\gamma)$ is the effective central charge and $S_0(\gamma)$ is the residual entropy.} After calculating the entanglement entropy for each trajectory, we average over different realizations, which we denote as \( \overline{S} \). In the subsequent sections, we study the behavior of \( \overline{S} \) in great detail for different frequency regimes. 

\begin{figure}
    \centering
    \includegraphics[width=0.48\textwidth]{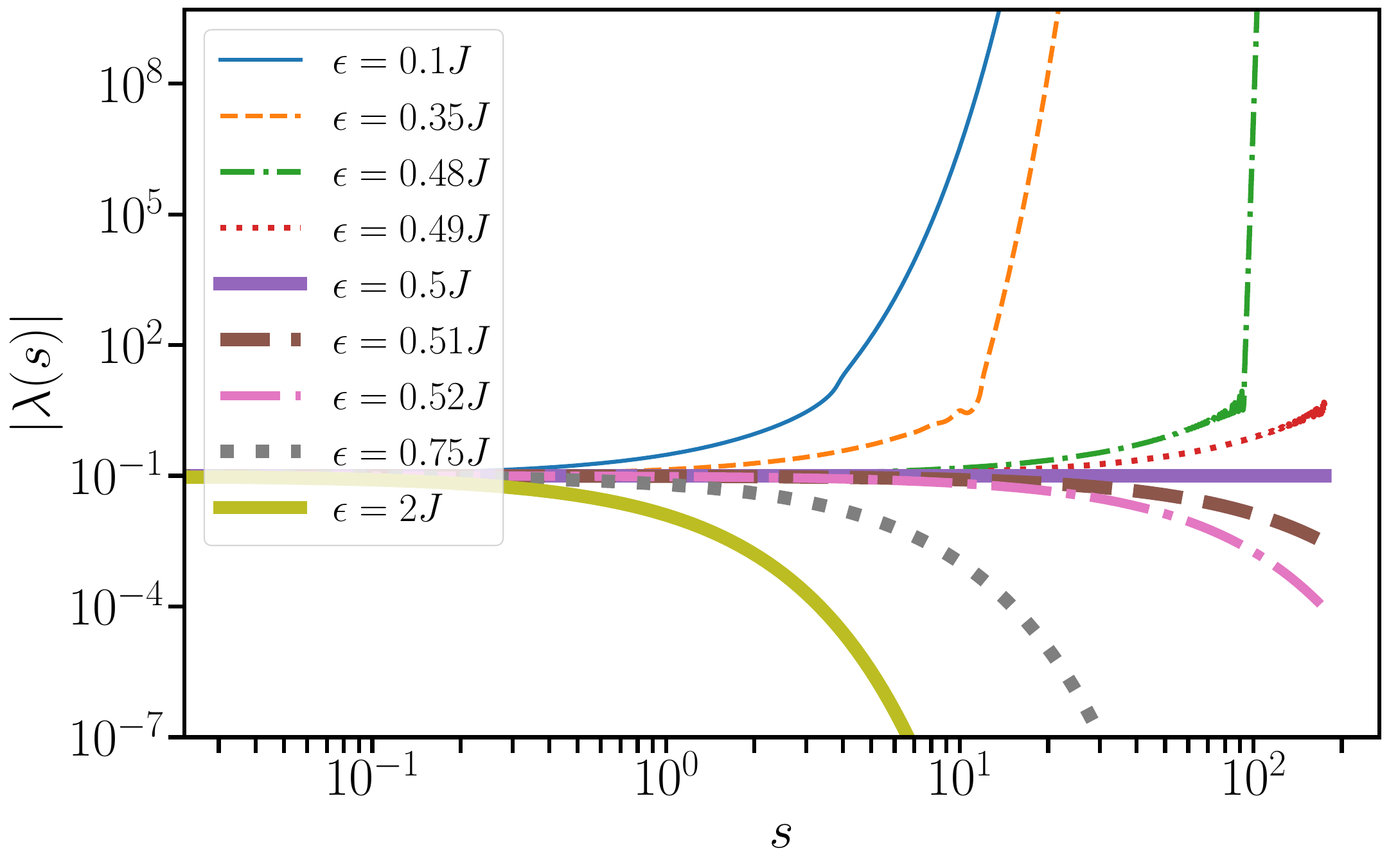}
    \caption{Flow of $\lambda$ for different $\epsilon $ values, with same initial condition $|\lambda=0.1|$ and $|K(\epsilon=2J)|=1.42\pi$. For $\epsilon=2J$, the flow of $\lambda$ is towards the Gaussian fixed point, which suggests a critical phase; on decreasing $\epsilon$, the flow of $\lambda$ approaches the strong coupling fixed point, which suggests area-law entanglement phase. In our calculation, we take $A=1$.}
    \label{fig_RG_flow}
\end{figure}
\subsection{High-frequency regime: Analytical prediction \label{analytic}}
It has been proposed that in the absence of the driving, the long-time dynamics of the continuously monitored fermionic Hamiltonian given by Eq.~\eqref{hamiltonian}  could be theoretically described by an effective bosonized non-Hermitian sine-Gordon model~ \cite{PhysRevX.11.041004},
\begin{eqnarray}
    H_{eff}=\frac{vJ}{2\pi}\int_x \bigg[(\partial_x\hat{\theta}_x)^2 + (1-\frac{2i\gamma}{vJ\pi})(\partial_x\hat{\phi}_x)^2\bigg] \nonumber \\+i\lambda\int_x\bigg[\cos(\sqrt{8}\hat{\phi}_x-1)\bigg].
    \label{main_eff}
\end{eqnarray}
Here, $v=\partial_k E(k)|_{k=k_F}, E(k)= 2\cos k$ for free theory and $k_F$ is fermi momentum, $\hat{\phi}_x$ is the bosonic field operator with its conjugate field $\hat{\theta}_x$ obeying the commutation relation  $[\partial_x{\hat{\theta}_x},\hat{\phi}_{x'}]=-i\pi\delta(x-x')$, and
$\lambda=\gamma m^2$, where $m$ is $O(1)$, depends on normal ordering prescription. It has been shown that in a long-time steady state, the system comes to a dark state, which is defined as the zero energy eigenstate of the $H_{eff}$, i.e., 
$ H_{eff}|\psi_D\rangle=0$ (or equivalently can be identified as an eigenstate of the evolution operator $U(T)=e^{-iH_{eff}T}$ with eigenvalues $1$, i.e., $U(T)|\psi_D\rangle=|\psi_D\rangle$). The steady-state entanglement entropy can be obtained from ~\cite{casini2009entanglement,calabrese2004entanglement},
\begin{equation}
    S(l)=\frac{1}{3} \langle \psi_D |\hat{\phi}_x\hat{\phi}_{x+l} |\psi_D\rangle.
    \label{ee_sine}
\end{equation}
\textcolor{black}{Note that this analysis has been questioned recently with the growing belief of the absence of measurement-induced phase transition in these models, and the analysis predicts a critical to area law phase transition in the no-drive scenario. However, since this study still captures the essential features of the measurement-induced transition  at least in the case of a finite-size system, and can be extended to the driven system—at least in the extreme high-frequency limit—it motivates us to carry out this analysis to develop some intuition about the driven case.}

In case of the square pulse driving protocol (as discussed in Sec.~\ref{model}), in a similar spirit, one can write down the effective Floquet evolution operator in terms of the non-Hermitian sine-Gordon Hamiltonian as
\begin{equation}
    U_{F}(T)=e^{-iH_{eff}^FT}=e^{-iH_{eff}^-T/2}e^{-iH_{eff}^+T/2},
\end{equation}
where,
\begin{eqnarray}
     H_{eff}^+=\frac{Jv}{2\pi}\int_x \bigg[(\partial_x\hat{\theta}_x)^2 + (1-\frac{2i\gamma}{Jv\pi})(\partial_x\hat{\phi}_x)^2\bigg] \nonumber \\+i\lambda\int_x\bigg[\cos(\sqrt{8}\hat{\phi}_x-1)\bigg].\nonumber
\end{eqnarray}
and
\begin{align}
 H_{eff}^-=\frac{(-J+\epsilon)v}{2\pi}\int_x \bigg[(\partial_x\hat{\theta}_x)^2 + &(1-\frac{2i\gamma}{(-J+\epsilon)v\pi})(\partial_x\hat{\phi}_x)^2\bigg] \nonumber \\&+i\lambda\int_x\bigg[\cos(\sqrt{8}\hat{\phi}_x-1)\bigg]. \nonumber
\end{align}
 \textcolor{black}{Construction of Floquet operator where a non-hermitian Hamiltonian governs the dynamics has been studied in Ref.~\cite{nonhermitian_floquet_1(2024), nonhermitian_floquet_2(2024)}}. In the high-frequency limit, the effective Floquet Hamiltonian $H_{eff}^F$ can be calculated using Magnus expansion as~\cite{bukov2015universal}, 
\begin{equation}
    H_{eff}^F= \frac{1}{2} (H_{eff}^- +  H_{eff}^+ )-\frac{iT}{4}[H_{eff}^-,H_{eff}^+]+O(T^2).
\end{equation}

\begin{figure}
    \centering
    \includegraphics[width=0.48\textwidth]{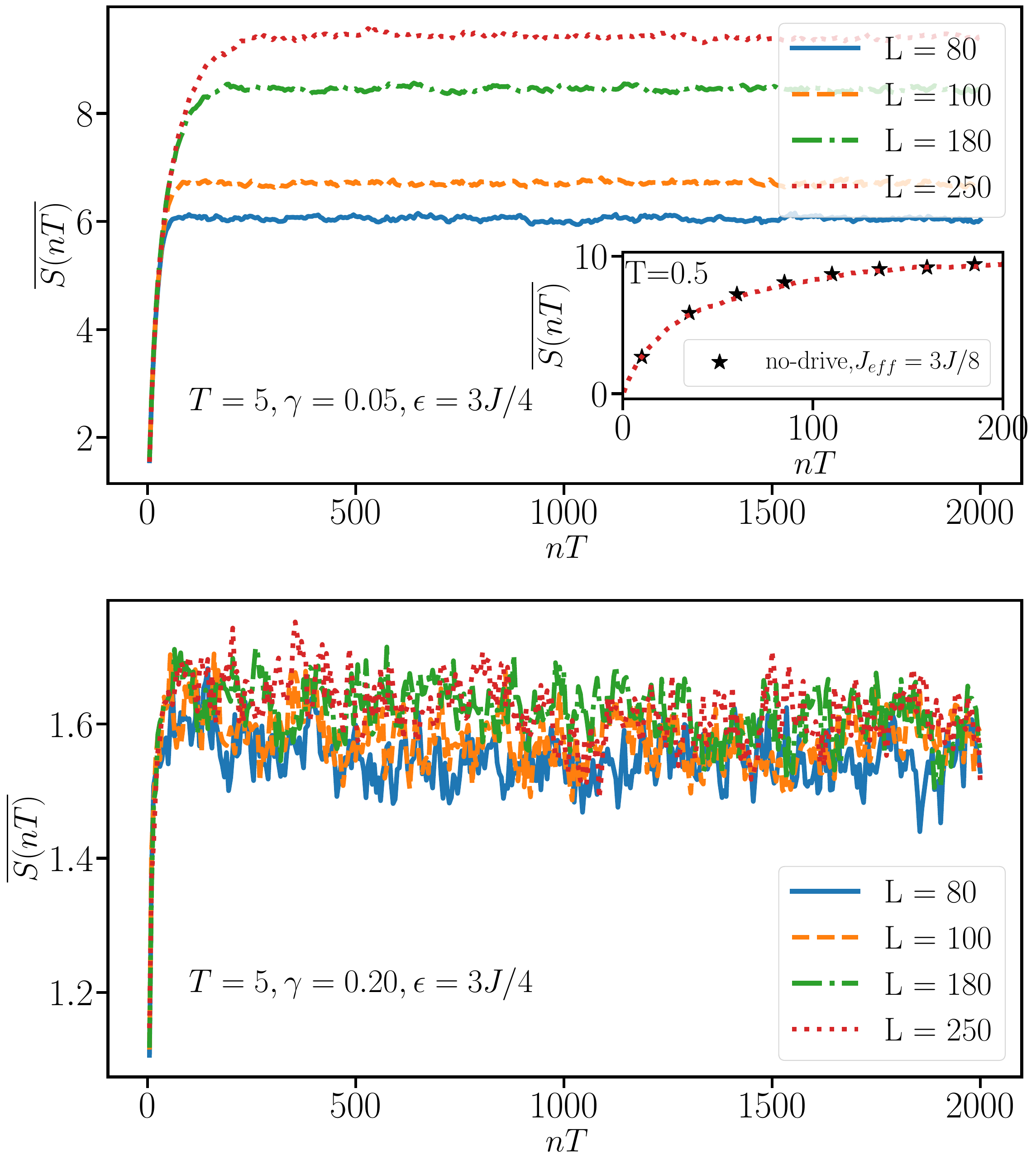}
    \caption{Stroboscopic growth of trajectory averaged entanglement entropy for $\gamma=0.05$ (upper panel) and $\gamma=0.20$ (lower panel). All plots are for  $\epsilon=3J/4$ and $T=5$. The upper panel inset shows that in the high-frequency regime $T=0.5$, the numerical results are in agreement with no-drive results for $J_{eff}=3J/8$. Inset data is for $\gamma=0.05$ and  $L=250$.}
    \label{fig_new_a}
\end{figure}

In the extremely high-frequency limit $(T<<1)$,  $H_{eff}^F$ can be approximated by the 1st term in the Magnus expansion as,
\begin{align}
     H_{eff}^F\simeq \bigg[&\frac{ v\epsilon}{4\pi}\int_x \bigg[(\partial_x\hat{\theta}_x)^2 + \eta^2(\partial_x\hat{\phi}_x)^2\bigg] \nonumber\\&+i\lambda\int_x\bigg[\cos(\sqrt{8}\hat{\phi}_x-1)\bigg],
     \label{final_H}
\end{align}
where $ \eta^2=1-\frac{4i\gamma}{\epsilon\pi v}$ . Hence, one would expect that 
in such a high-frequency limit, the steady state of the driven continuously monitored system, as discussed in Sec.~\ref{model}, can be described by the dark state of the non-Hermitian sine-Gordon Hamiltonian Eq.~\eqref{final_H} and steady state entanglement entropy can be obtained using Eq.~\eqref{ee_sine}, $|\psi_D\rangle$ correspond to the zero energy eigenstate of the Hamiltonian Eq.~\eqref{final_H}.  
\textcolor{black}{Moreover, this also suggests that the entanglement dynamics in the high-frequency limit can effectively be mimicked by the continuously monitored no-drive dynamics of tight-binding Hamiltonian with effective hopping amplitude $J_{eff}=\epsilon/2$. We checked numerically in Section. \ref{numerics} in Fig. \ref{fig_new_a} upper panel inset, taking a very small time period $T=0.5$ for a given $\epsilon $ and the dynamics is completely matching with $J=\epsilon/2$ model.} It also automatically implies if in the no drive case, i.e., $\epsilon=2J$, the transition point is at $\gamma_c=\gamma_{0}$ (which is predicted by our RG analysis), for $\epsilon < 2J$ (the drive is introduced), the transition point will be $\gamma_c\simeq\epsilon\gamma_{0}/2J$, decays to zero linearly with decreasing $\epsilon$. 

In the case of the $\epsilon>0$ and $\lambda=0$, the Hamiltonian Eq.~\eqref{final_H} has a Gaussian fixed point, which implies the logarithmic scaling of the entanglement entropy with sub-system size will prevail~\cite{PhysRevX.11.041004}, i.e., 
\begin{equation}
    S=\frac{1}{3} c({\gamma})\ln(l),
\end{equation}
where $c({\gamma})$ is referred to as the effective central charge. We refer to this phase as a critical phase. One needs to find out the renormalization flow equations to study the effect of the nonlinearity $\lambda$ and investigate the fate of the Gaussian fixed point.  The Hamiltonian Eq.~\eqref{final_H} can be brought in its action form using complex wick rotation $(x,t)\to ({\tilde{\epsilon}}^{1/2}{\eta}^{1/2}x, i\tilde{\epsilon}^{-1/2}{\eta}^{-1/2}t)$, and the action $\mathcal{F}$ reads as, 
\begin{equation}
    \mathcal{F}=\int_{X}\frac{K}{16\pi}(\grad \hat{\phi}_X)^2 + i{\lambda }\cos(\hat{\phi}_X),
\end{equation}
where $(\grad \phi_x)^2= (\partial_x\hat{\phi}_X)^2 + (\partial_t\hat{\phi}_X)^2, X \in(x,t) $, $\tilde{\epsilon}=v\epsilon/2\pi$, and $K^2=\eta^2$. We use a similar renormalization group (RG) prescription as discussed in Ref.~\cite{PhysRevX.11.041004,amit1980renormalisation}, first, decompose the fields into short-range modes that correspond to momentum $\frac{\Lambda}{\zeta}<|k|<\Lambda$ and the long-range modes that correspond to  $|k|<\frac{\Lambda}{\zeta}$ ($\Lambda$ is a
short-distance cutoff, and $\zeta=e^s$, and $s$ is the rescaling parameter controlling
the renormalization group flow), and finally, integrate out the short-range modes.  One can write down renormalization group flow equations perturbatively in ${\lambda}$ as (note these equations are identical as obtained in Ref.~\cite{PhysRevX.11.041004}) ,
\begin{align}
    &\partial_s \lambda=(2-\frac{8\pi}{K})\lambda\nonumber\\
    &\partial_s K=-\lambda^2 A,
    \label{rg flow}
\end{align}
where $A$ is a positive number of order $O(1)$ determined by the propagator of the Gaussian theory. If increasing the RG iteration step $\lambda$ grows, it signifies that the non-linearity is a relevant perturbation; on the other hand, if $\lambda$ decays, that will make the perturbation irrelevant, and the physics will still be governed by the Gaussian fixed point, implying the survival of the critical phase. 
In Fig.~\ref{fig_RG_flow} we numerically solve the Eq.~\eqref{rg flow} with initial condition $|\lambda(s=0)|=0.1$ and $K(s=0,\epsilon)=\sqrt{\frac{2J}{\epsilon}{K^2(s=0,\epsilon=2J)}+1-\frac{2J}{\epsilon}}$, with $|K(s=0,\epsilon=2J)|=1.42\pi$. \textcolor{black}{These initial conditions are not special apart from the fact that for such initial conditions in the  $\epsilon=2J$ limit (no-drive scenario),  $\lambda$ always decreases with increasing $s$, signifying RG flow towards the Gaussian fixed point, implying the critical phase. The question we ask is: if we change $\epsilon$ (an indirect way to introduce drive), how will $\lambda$ change with RG steps? 
We see,  on decreasing $\epsilon$, the $\lambda$ starts growing with $s$, which makes the non-linear $\lambda$ term relevant, while for $\epsilon=0.5J$ we get that the flow is marginal for our chosen initial condition. Also, we have checked that with increasing the initial value of the parameter $|K(\epsilon=2J)|$, the marginal $\epsilon$ value also increases.} It signifies that physics is no longer governed by the Gaussian fixed point; instead, it is governed by a strong coupling fixed point, which is responsible for the area-law entanglement phase.  This study also shows that decreasing $\epsilon$ promotes the area-law entanglement phase, and if one starts with an initial condition such that the no-drive scenario shows a critical phase, with decreasing $\epsilon$, one should be able to probe the critical-area law phase transition~\cite{diehl_prl_2021}.

Next, we try to understand the effect of $\epsilon$ on the critical phase, more precisely, how the correlation or entanglement entropy depends on our drive protocol. We just focus on the $\lambda=0$ limit of the Hamiltonian ~\eqref{final_H}, which can be expressed in the momentum space representation   as,
\begin{align*}
    H^{F}_{eff}(\lambda=0)=\frac{v\epsilon}{4\pi}\int_q q^2\bigg[\hat{\theta_q} \hat{\theta}_{-q} + \tilde{\eta}^2 \hat{\phi}_q \hat{\phi}_{-q}\bigg].
\end{align*}
It has been argued that this Hamiltonian has a unique dark state~\cite{PhysRevX.11.041004}, and the correlation function in this dark state is given by, 
\begin{align*}
    \langle \psi_D | \hat{\phi}_q \hat{\phi}_{-q}|\psi_D\rangle=\frac{v^{3/2}\sqrt{\pi\epsilon}}{8\gamma|q|}\sqrt{2\sqrt{16\gamma^2+{(v\pi\epsilon)}^2}-2v\pi\epsilon}.
\end{align*}
It is straightforward to check that for a fixed non-zero $\gamma$, in the limit $\epsilon \to 0$, this correlation tends to zero, and the effective central charge $c(\gamma)$ (note that the leading order term in the entanglement entropy $S=\frac{c(\gamma)}{3}\ln l$) will also approach zero as $\sqrt{\frac{\epsilon}{\gamma}}$. This implies that in the case of $\epsilon=0$,  for any finite $\gamma$, one would expect to observe the area-law phase. In a nutshell \textcolor{black}{in case of finite size systems}, we make four main predictions based on this RG analysis about the high-frequency limit drive: 1) the entanglement dynamics can be mimicked by the dynamics of no-drive continuously monitored free-fermionic chain with the hopping amplitude  $J_{eff}=\epsilon/2$, 2) If one starts in critical-phase in the no-drive regime, i.e. $\epsilon=2J$, with decreasing $\epsilon$, one can be able to probe critical to area-law phase transition, 3) $\epsilon = 0$ limit will not display entanglement transition, it will always show the area-law entanglement for any non-zero finite measurement strength, 4) the transition point $\gamma_c$ is expected to decay to zero linearly with decreasing $\epsilon$ in the high-frequency region.  

In this context, we would like to mention that very recently, a modified version of the RG analysis on sine-Gordon type models has been proposed after refining their previous analysis, in Ref.~\cite{arealaw_2_muller2025monitored}, which shows the critical to area law transition point shifts towards zero in the case of a non-interacting monitored system in the absence of drive. However, the symmetries of these continuum models are not exactly identical to our lattice setup. In monitored systems, energy conservation is broken, and measurements can populate fermions across the entire band. Hence, there is no guarantee that any such models (that include Hamiltonian Eq.~\eqref{main_eff} too) can describe the dynamics of monitored fermions on a tight-binding lattice. Also, the prediction of Ref.~\cite{arealaw_2_muller2025monitored} of a logarithmic to area law phase transition in the interacting case, contradicts the other recent works~\cite{Interacting_NLSM_1,fazio.prbl.24, guo2024field}, where they observed a volume-law to area-law phase transition with increasing measurement rate. In our study, the RG scheme we used has helped us build an intuition about measurement-induced phase transitions in the presence of drive, which we will test in the subsequent section with finite-sized numerics. 


\begin{figure}
    \centering
    \includegraphics[width=0.48\textwidth]{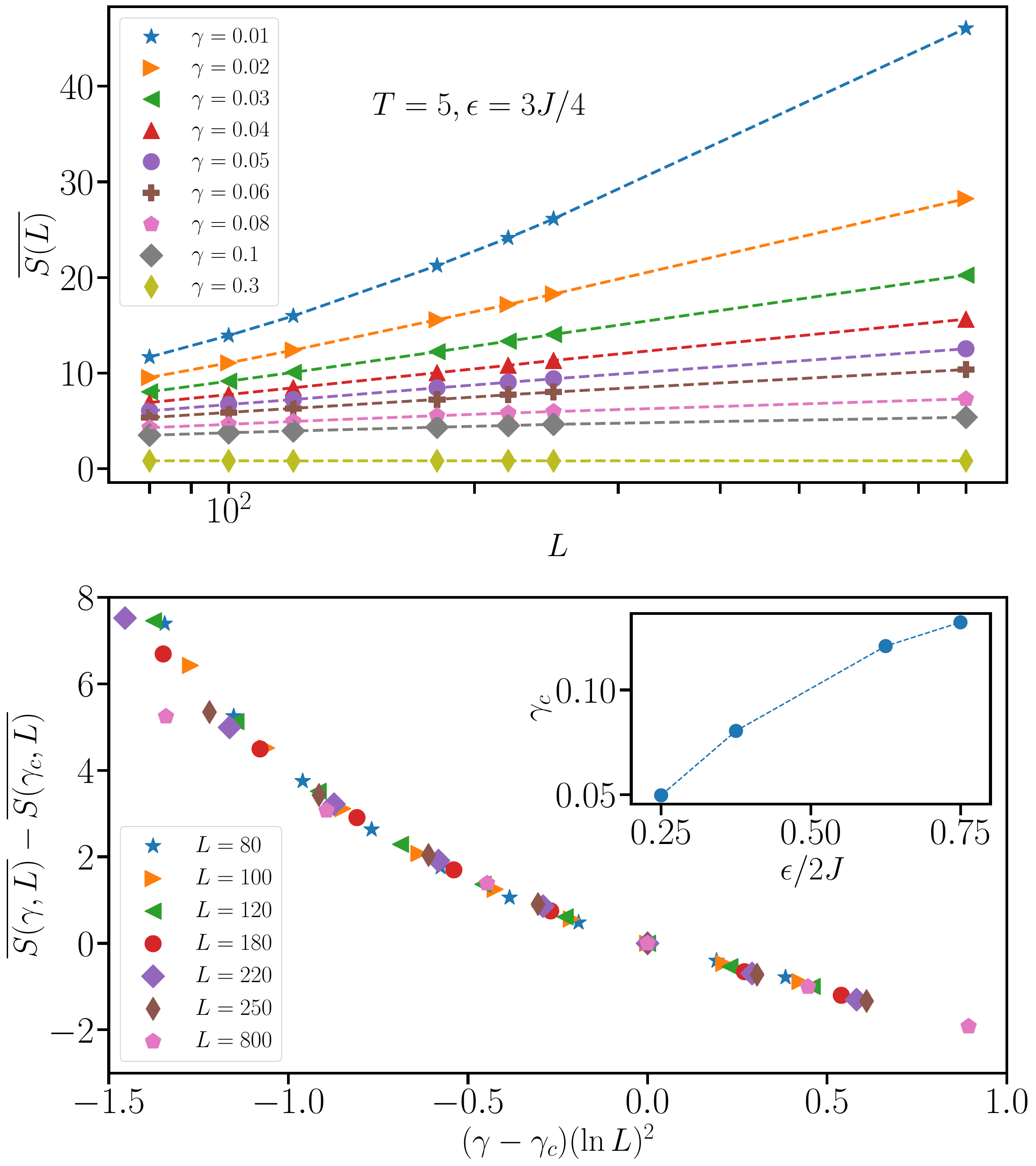}
    \caption{Upper Panel: System size scaling of steady state trajectory averaged entanglement entropy $\overline{S(L)}$ for different $\gamma$. Lower Panel: Data collapse of entanglement entropy for different $L$ and $\gamma$ assuming the BKT type scaling ansatz with critical measurement strength $\gamma_c=0.08$. The results are for $T=5$ and $\epsilon=3J/4$. The inset in the lower panel shows the variation of the  $\gamma_c$  (obtained from the data-collapse) for different $\epsilon$. }
    \label{fig_new_b}
\end{figure}
 \begin{figure}
    \centering
    \includegraphics[width=0.49\textwidth]{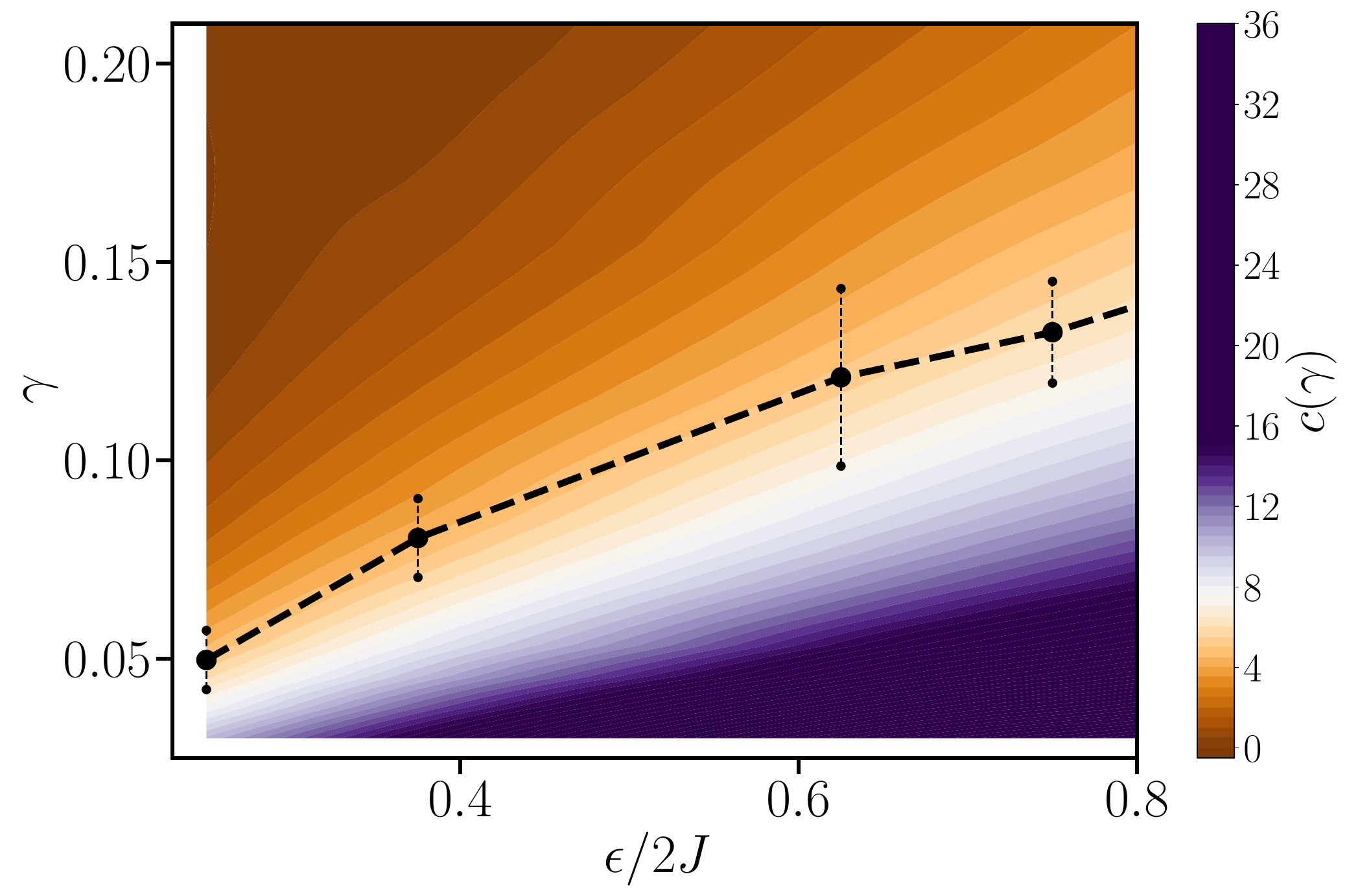}
    \caption{Color variation corresponds to the different values of the effective central charge $c(\gamma)$ for $T=5$. Black Dots represent $\gamma_c$ for finite-size system.}
    \label{fig_contour}
\end{figure}

 \subsection{Numerical results\label{numerics}}
 In this section, we perform a numerical experiment on a finite-size lattice with periodic boundary conditions using the methodology described in Sec.~\ref{model}. \textcolor{black}{We follow the following strategies; first, we will pretend the BKT transition takes place for the driven system, and compute critical measurement strength $\gamma_c$ using appropriate BKT ansatz, then we compute the local effective central charge~\eqref{c_eff_eq} as proposed in Ref.~\cite{starchl2024generalized}, and investigate whether for $\gamma <  \gamma_c$ (obtained from scaling analysis) the local effective central charge start decreasing with increasing system size. If we don't find such a signature, then we would conclude that the critical phase is stable. }

As a beginning, we study the time evolution of the entanglement entropy, i.e., averaged over many different trajectories, starting from the initial zero entanglement product state for different values of  $\epsilon$ and measurement strengths $\gamma,$ for $T=5$. Figure.~\ref{fig_new_a} shows the dynamics of $\overline{S(nT)}$ for $\epsilon=3J/4$. The upper panel shows that for $\gamma=0.05$, the saturation value of the entanglement entropy increases with system size; on the other hand, the lower panel shows that for $\gamma=0.2$, the steady state value of the average entanglement entropy hardly changes with system size. 
 This is a signature that, with increasing measurement strength, an entanglement transition may occur, where for small values of $\gamma$, the steady-state entanglement increases with $L$, and for large $\gamma$, it does not scale with $L$. 
 The inset of the upper panel shows the entanglement dynamics in the high-frequency regime $T=0.5$ for $\gamma=0.05$ and $L=250$ (red-dotted line).
  As predicted in the previous section, we find that, indeed, for $T=0.5$, the entanglement dynamics can be mimicked by the continuously monitored no-drive scenario in a tight-binding lattice with an effective hopping strength $J_{eff}=\epsilon/2$, which is denoted by points in the inset figure.  \textcolor{black}{Given, the results for $T<1$ are almost indistinguishable from the no-drive results, at least for our accessible system size, hence, we consider $T=5 >1$ for most of our numerics. The no-drive results are separately presented in the Appendix.~\ref{nodrive_central_charge}}

  \textcolor{black}{Next, we want to see how entanglement entropy scales with the subsystem sizes, and want to predict the critical $\gamma_c$(if it exists) at which an entanglement phase transition happens. For this, we opt for two methods. In one case (methodology 1), we vary the total system size $L$ and always take the subsystem size as $l=L/2$. In the other method (Methodology 2), we take a large system size $L$, and calculate the entanglement entropy for different subsystem sizes $l$.}
  \paragraph*{Methodology 1:}
  We check the system size scaling of steady-state entanglement entropy for $\epsilon= 3J/4$ in Fig.~\ref{fig_new_b} (upper panel) for different $\gamma$ values. It shows evidence of a logarithmic-area law entanglement phase transition. 
  To reconfirm that for small $\gamma$, the scaling is logarithmic (not volume law), we also perform an F-test (see Appendix.~\ref{f-test}).
 \textcolor{black}{A plethora of studies in the last few years claim that, in the absence of driving, the measurement-induced critical-to-area-law phase transition with increasing $\gamma$ belongs to the BKT universality class~\cite{areejit_prb, diehl_prl_2021,PhysRevLett.128.010605,PhysRevLett.128.010603}. However, the existence of such a transition in the thermodynamic limit has been questioned very recently; it has been proposed that even for small $\gamma$, the steady-state entanglement entropy will show area-law scaling~\cite{mirlin.2023}. At least for numerically accessible system sizes and for $\epsilon=3J/4$, we do not observe such signatures; the data mostly support the logarithmic scaling of the steady-state entanglement entropy with system size. Hence, we fit our data with the BKT scaling ansatz.}
  The system size dependence of the entanglement entropy in the BKT universality class obeys the following scaling form in the vicinity of the critical $\gamma_c$~\cite{harada1997universal},
 \begin{equation}
     \bar{S}(L/2,L,\gamma)-\bar{S}(L/2,L,\gamma_c)=F[(\gamma-\gamma_c)(\ln L)^2].
     \label{entangle_data_collapse}
 \end{equation}
Given that the steady entanglement entropy growth is logarithmic in a small $\gamma$ regime, we also use a similar BKT scaling ansatz to obtain the critical $\gamma_c$ values for different $\epsilon$ and $T=5$. Figure.~\ref{fig_new_b} (lower panel) shows the data collapse of steady state entanglement entropy for different $L$ and $\gamma$, which allows us to evaluate the critical $\gamma_c$. We use the cost function minimization technique to obtain the $\gamma_c$ (see Appendix.~\ref{cost}). The inset shows the critical $\gamma_c$ values for different $\epsilon$. Our results suggest the critical $\gamma_c$  decreases with decreasing $\epsilon$ and approaches zero in the $\epsilon\to 0$ limit, as also predicted by our previous RG study for the high-frequency limit. \textcolor{black}{
Figure.~\ref{fig_contour} shows the contour plot of the effective central charge $c(\gamma)$ (which is obtained by fitting the steady-state entanglement data with $L$) for different values of $\gamma$ and $\epsilon$, and dots correspond to the $\gamma_c$.  We discuss $\epsilon/2J=1$ (no-drive) results separately in the Appendix \ref{nodrive_central_charge}.}
 \textcolor{black}{However, obtaining $\gamma_c$ accurately for large $T$ is extremely difficult. It turns out that for a given large $T$, there are system sizes $L < L_1$, for which the steady-state entanglement entropy is the same as the no-drive scenario with $J_{\text{eff}} = |J - \epsilon|$. The effect of the drive kicks in for $L > L_1$, where $L_1$ increases with $T$. This makes it extremely difficult to access the system size regime required to obtain the proper scaling. However, we do a comparative study of the results for different $T$ in Sec.~\ref{T_dep} later.}
Moreover, the data for $T=5$ also suggests that for a symmetric drive with zero means, i.e., for  $\epsilon=0$, $\gamma_c\to 0$, implies no transition, always an area-law phase. 
Though our analytical results suggest the same for the high-frequency regime, it is not at all obvious that this phenomenon persists even for large $T$. Especially given that there are examples where disorder-induced localization-delocalization transition has been observed in the large $T$ limit for symmetric drive mean around zero~\cite{PhysRevB.103.184309, PhysRevB.105.024301}. Hence, in Section \ref{e0}, we investigate the $\epsilon=0$ limit even more carefully. 

\paragraph*{Methodology 2:}\label{Alternative analysis}
\begin{figure}
    \centering
    \includegraphics[width=0.48\textwidth]{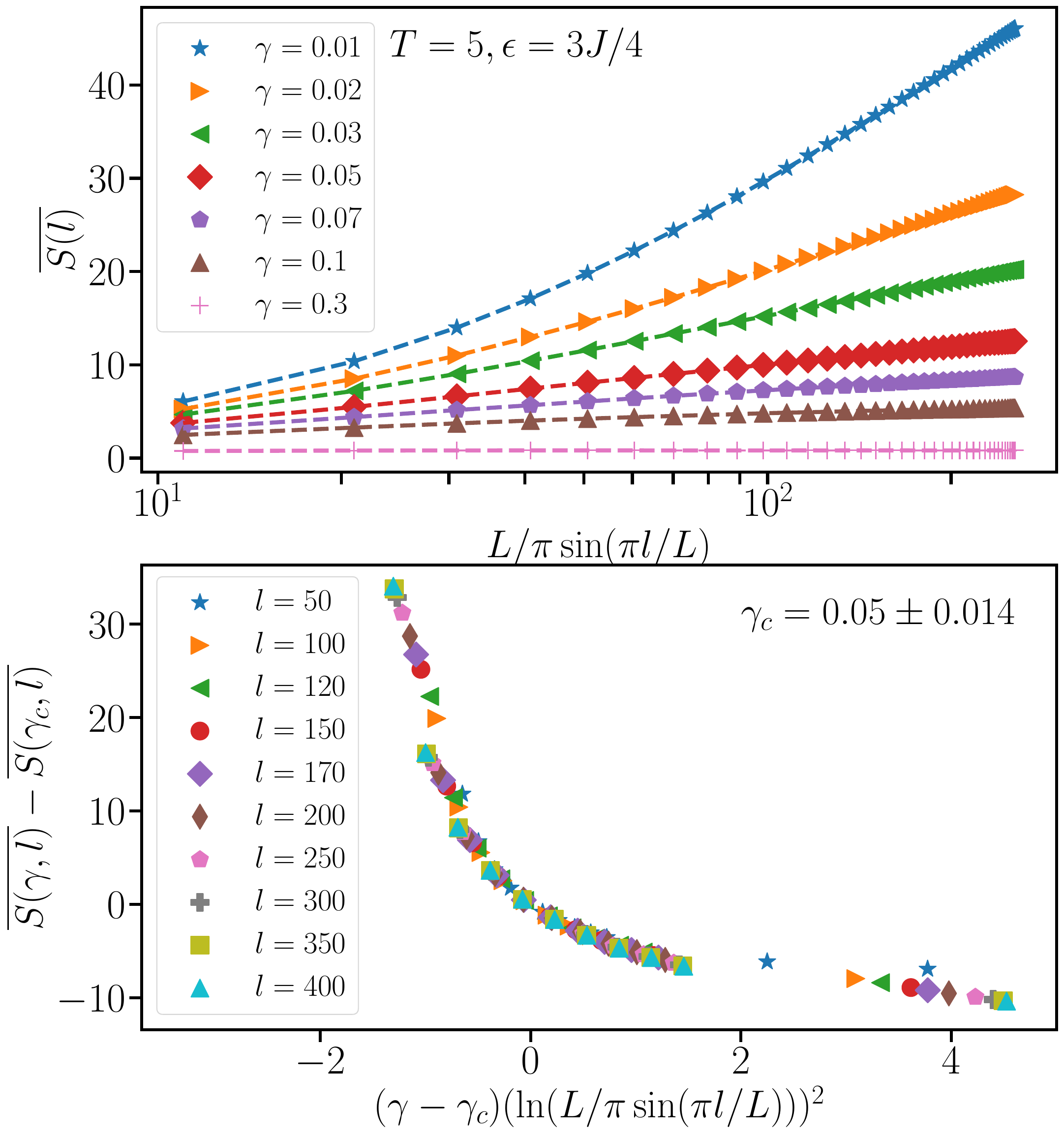}
    \caption{$T=5$ data of entanglement entropy scaling with subsystem size $l$(upper panel) and finite-size data collapse of entanglement entropy for different subsystem size(lower panel) for $L=800$.}
    \label{diff_bipart_data_collapse}
\end{figure}
\textcolor{black}{Previously, to compute the steady state entanglement, we had restricted ourselves to the sub-system size  $l=L/2$ and then increased $L$ to obtain finite size scaling.   
In this section, we present the entanglement entropy data for which we fix the total system size to $L\simeq 10^3$, and vary the sub-system size $l$. Once again,  the entanglement entropy can be computed using Eq.~\eqref{ent_la}, and 
 we present the data of entanglement growth with subsystem size in Fig. ~\ref{diff_bipart_data_collapse} upper panel for $T=5$ and $L=800$ for $\epsilon=3J/4$. To obtain the critical $\gamma_c$, we use the same cost function minimization technique satisfying the scaling law in the vicinity of the critical phase as,
\begin{equation}
     \overline{S(l,\gamma)}-\overline{S(l,\gamma_c)}=F[(\gamma-\gamma_c)(\ln (\frac{L}{\pi} \sin{\frac{\pi l}{L}}))^2].
     \label{new_entangle_data_collapse}
 \end{equation}
 Fig. ~\ref{diff_bipart_data_collapse} lower panel shows the data collapse of the entanglement entropy for different subsystem size $l$, with $\gamma_c= 0.05 \pm 0.014$ calculated using the cost function minimization presented in the Appendix. ~\ref{cost} in Fig.~\ref{cost_new}.} 
 
 \paragraph*{Local effective central charge:}
 \begin{figure}[!ht]
    \centering
    \includegraphics[width=0.48\textwidth]{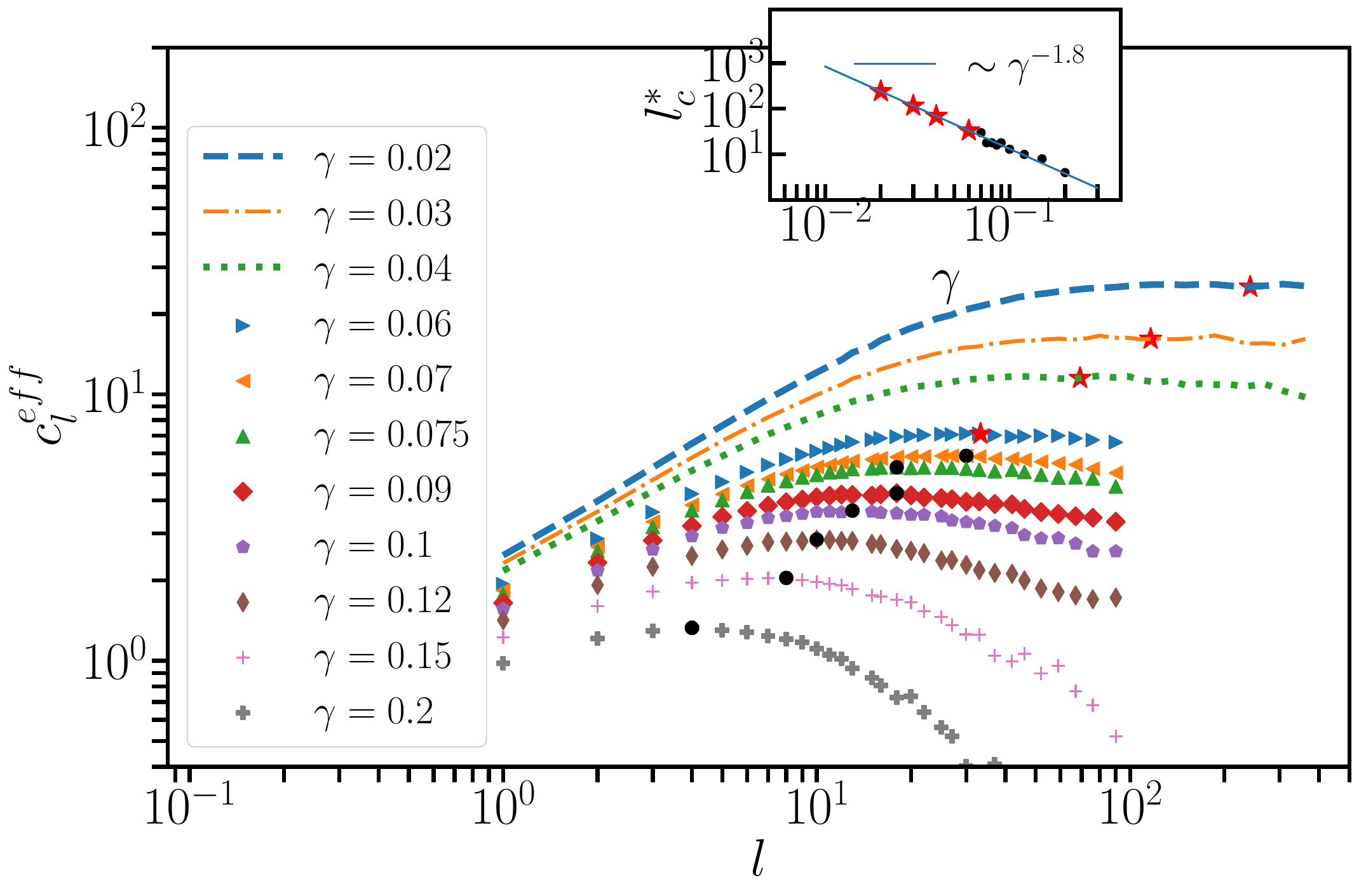}
    \caption{Effective central charge $c^{eff}_l$ vs. $l$ plots for different $\gamma$ values for $T=5, \epsilon= 3J/4$. Black circular points represent the maximum of the central charge and red star points represent the position of $l^*_c$ for which $c^{eff}_l$ is the maximum estimated from the fitting. The inset shows the best fit, which gives $l^*_c \sim \gamma^{-1.8}$. The maximum system size is taken as $L=1004$. }
    \label{C_eff_T_5}
\end{figure}
 \textcolor{black}{
 Here we compute subsystem-dependent effective central charge and analyze our data as was done in a recent Ref. ~\cite{arealaw_1_starchl2024generalized} for the no-drive scenario.
 We calculate the effective central charge defined as
\begin{equation}
   c^{eff}_l=3 \frac{\overline{S}_{l'}-\overline{S}_l}{\ln{\frac{\tilde{l'}}{\tilde{l}}}},
   \label{c_eff_eq}
\end{equation}
where, $\tilde{l}=\frac{L}{\pi} \sin(\pi l/L)$ is the cord length and $l$ is the size of the subsystem varies from $1$ to $L/2$. For logarithmic growth of the entanglement entropy, this measure should be constant; for volume-law scaling,
it should scale as $c^{eff}_l \sim l$, while area-law behavior leads to $c^{eff}_l=0$. To calculate $c^{eff}_l$ numerically, we choose
$N_l$ values of  $l_i$ with $l_1 = 1$ and $l_{N_l} = L/2$ such that the successive ratio of $\frac{\tilde{l'}}{\tilde{l}}$
is approximately constant. In Ref.~\cite{arealaw_1_starchl2024generalized}, it was shown that the critical phase is bounded from below by $l \sim \gamma^{-1}$ and from above by $l \sim \gamma^{-2}$. If the effective central charge increases with system size and attains a maximum, then it slowly starts decreasing, which indicates a crossover to area law.
In Fig.~\ref{C_eff_T_5} we present the effective central charge $c^{eff}_l$ data for the driving period $T=5$. We see a very clear evidence of bending of $c^{eff}_l$ after a critical $l^*_c$ for $\gamma>\gamma_c=0.05\pm 0.014$ (black circles represent the maximum). We fit those $l^*_c$ and obtain the scaling $l^*_c \sim \gamma^{-1.8}$. Following the fitting results, we estimate the $l^*_c$ values for $\gamma \le \gamma_c$, which is represented by the red star points. Then we compare these $l^*_c$ with actual simulated data for system size $1004$, and we get only a very mild signature of bending for $\gamma=0.04$ (which is still in the regime of the critical $\gamma_c$ with our estimated error-bar), but for $\gamma=0.03, 0.02$ (they are $<\gamma_c$) the bending is not at all clearly evident for the mentioned system size. We can interpret this in two different ways, either $l^{*}_c$ scales faster than power-law $~\gamma^{-1.8}$ in the case of small $\gamma$ limit, so that the $l^*_c$ (beyond which $c^{eff}_l$ will show the bending) is very large for small $\gamma$ values, which is not accessible numerically, or there exist a critical phase for $\gamma <\gamma_c$, in that case data will never bend. Results in the absence of periodic drive are also presented in the Appendix.~\ref{nodrive_central_charge} for our protocol for comparison. \textcolor{black}{We have also presented the equal time connected correlation function calculations in Appendix~\ref {corr func} as another tool to support the entanglement entropy data for $T=5$ for finite size systems.}
}

\paragraph*{Dependancy of steady state entanglement entropy on driving frequency:}\label{T_dep}

\begin{figure}[!ht]
    \centering
    \includegraphics[width=0.48\textwidth]{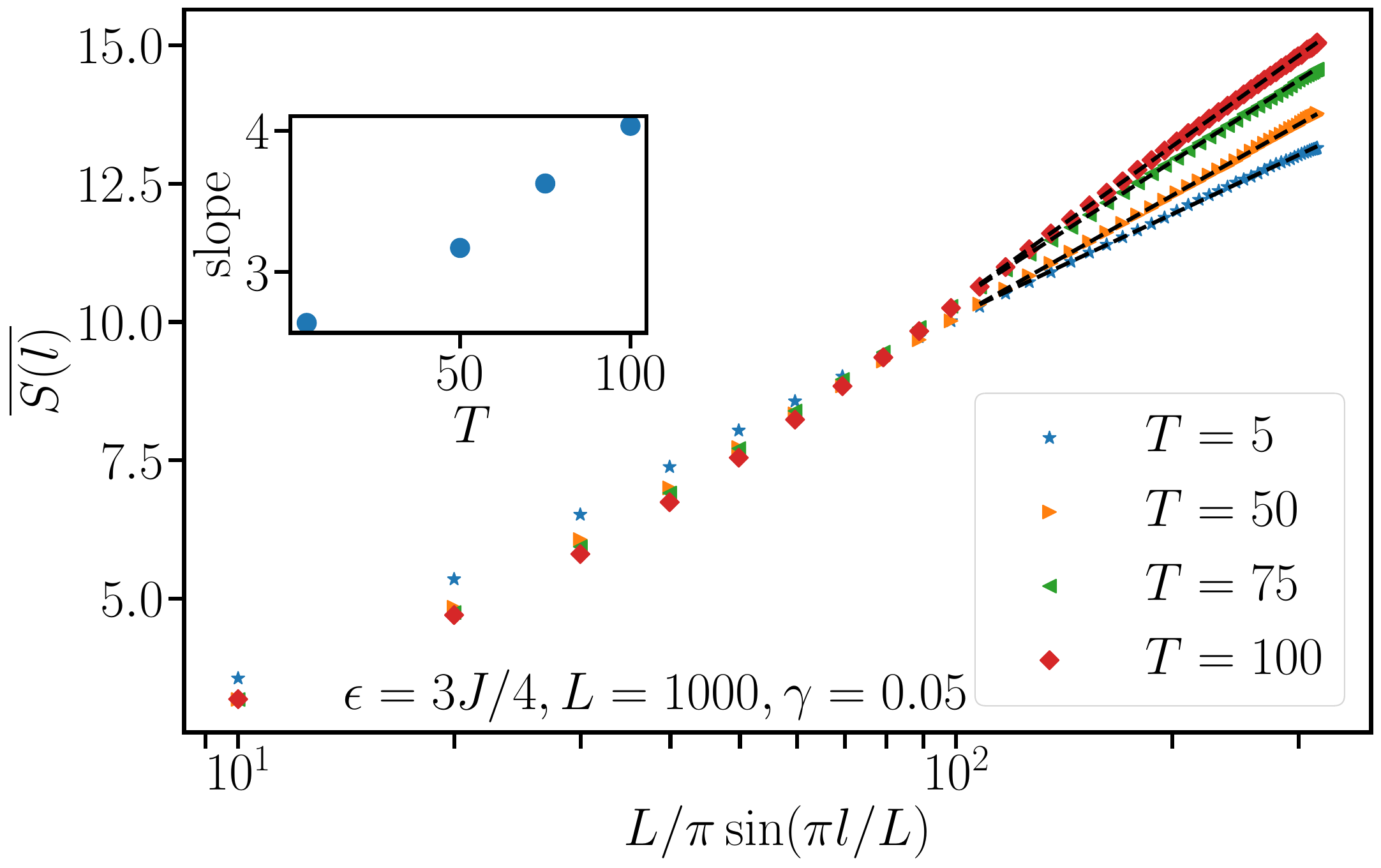}

    \caption{ Comparison of entanglement entropy growth for different driving periods $T$ with system size $L=1000$ for $\gamma=0.05$. The inset shows that the slope of the entanglement growth is increasing with increasing $T$ for $\epsilon=3J/4$. }
    \label{entanglement_mean}
\end{figure}
\begin{figure}
  \centering
    \includegraphics[width=0.48\textwidth]{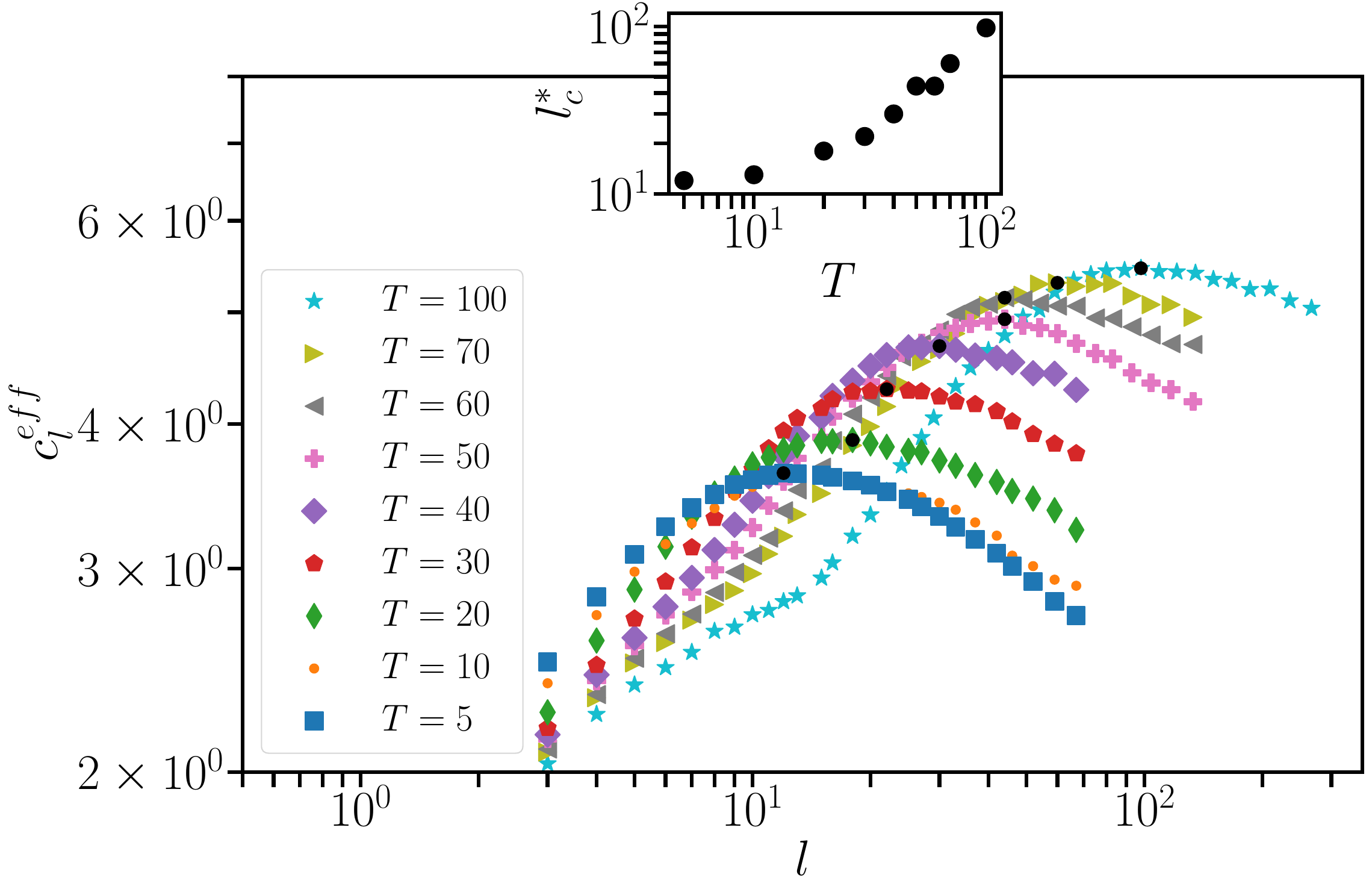}
    \caption{Effective central charge $c^{eff}_l$ vs. $l$ plots for different $T$ values for $\gamma=0.1, \epsilon= 3J/4$. Black circular points represent the maximum of the central charge. The inset shows that $l_c^*$ is increasing with increasing $T$. The maximum system size is taken as $L=1004$. }
\label{figA1}
   \end{figure}
\begin{figure}
  \centering
    \includegraphics[width=0.48\textwidth]{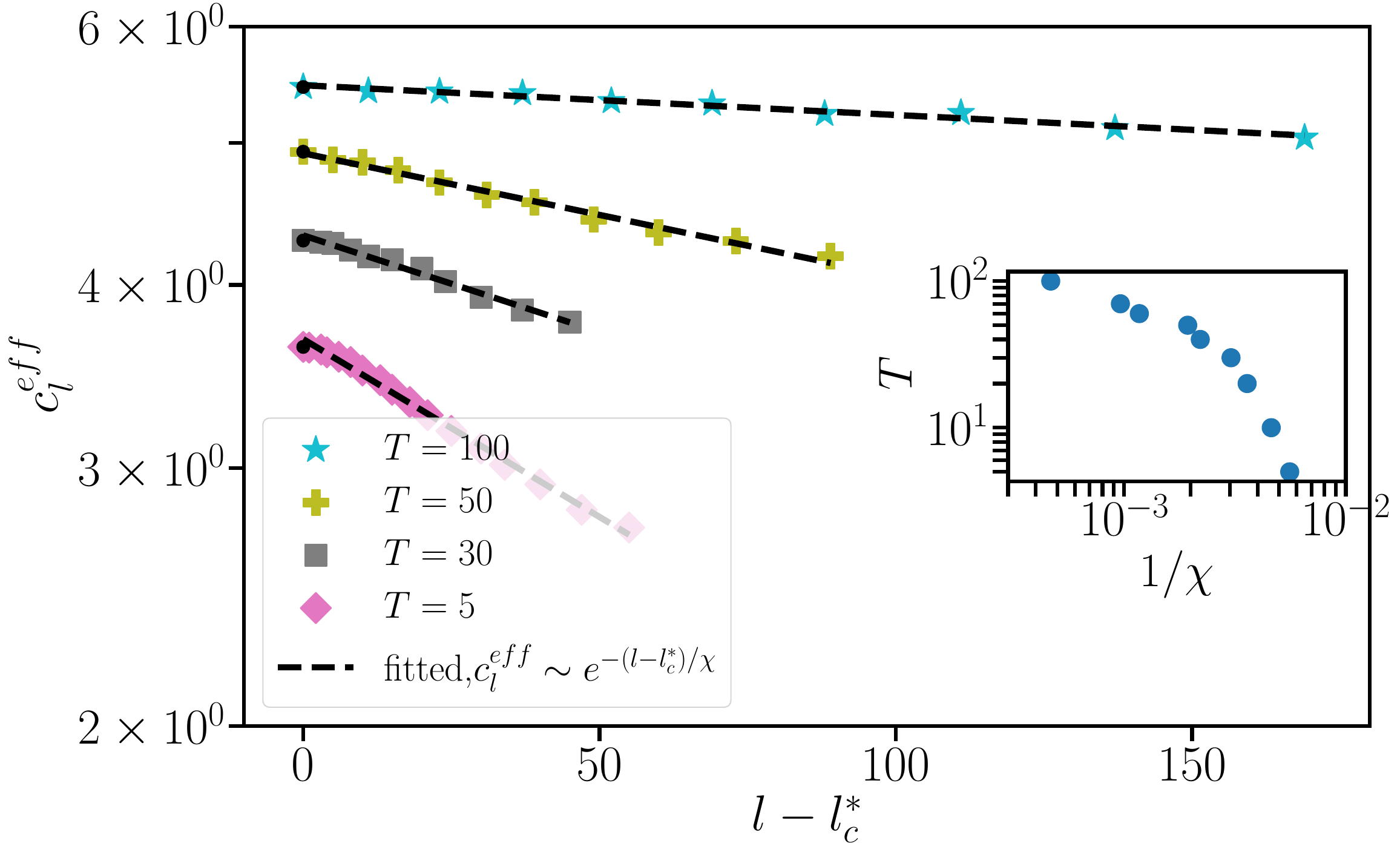}
    \caption{$c_l^{eff}$ vs. $l-l_c^*$ plot with $l> l_c^*$ for different $T$ values. $l_c^*$ represents the value of $l$ for which $c_l^{eff}$ shows a peak. Black dashed lines shows the fitting having the functional form $c_l^{eff} \sim e^{-(l-l_c^*)/\chi}$. The inset shows that $\chi$ is increasing with increasing $T$. The maximum system size is taken as $L=1004$.  }
  \label{figA2}
   \end{figure}

\textcolor{black}{Next, we investigate the effect of $T$ in the steady state entanglement scaling with subsystem size for a fixed $\epsilon=3J/4$. As mentioned earlier, as we increase $T$, it becomes even harder to compute $\gamma_c$; hence, we instead study the steady state entanglement scaling vs subsystem size for a given $\gamma$ and for different $T$.  Figure~\ref {entanglement_mean} shows that in the large $l$ limit, the slope of the trajectory-averaged long-time entanglement entropy increases with increasing driving period $T$. 
The inset shows the slope of the entanglement entropy for different $T$, where we calculate the slope by fitting the entanglement entropy data in the large $l$ limit for $\gamma=0.05$ and $\epsilon=3J/4$, for system size $L=1000$. 
If the BKT transition survives in the thermodynamics limit for $T=5$ (which was one of the potential outcomes of our previous analysis), we can now infer from these results that increasing $T$ favors the critical phase more, and if the BKT transition survives, then the critical measurement strength $\gamma_c$  will increase with increasing $T$.}

\textcolor{black}{Next, to understand the effect of the driving frequency on measurement-induced transition in more detail, we investigate a scenario where we fix $\gamma=0.1$ for which $T=5$ data of effective local central charge clearly shows bending in the large $l$ limit (suggesting area law scaling), and then increase $T$. We observe two things (see Fig.~\ref{figA1}): 1) the $l^*_c$ (the sub-system size for which the $c^{eff}_l=c_{eff}^{max}$ is maximum) increases with $T$ (this $c_{eff}^{max}$ is a measure of the maximum growth rate of the steady state entanglement
with sub-system size, which we have plotted in Fig.~\ref{fig0} along with the schematic of our protocol), 2) the bending of the $c^{eff}_l$ data for $l>l^*_c$ becomes much slower with increasing $T$. To quantify the bending rate of the data, we plot $c^{eff}_l$ vs $(l-l_c^*)$ for $l>l^*_c$ in Fig.~\ref{figA2}, and find that it decreases exponentially with $l$. We fit the data with the function $c_l^{eff} \sim e^{-(l-l_c^*)/\chi}$ for $l>l_c^*$, where a length scale $\chi$ is a measure of how much the data has bend, and $\chi\to \infty$ limit should correspond to the absence of the bending of  $c_l^{eff}$ data, and that implies $c_l^{eff}$ is constant beyond $l=l^*_c$. This will be a signature of the critical phase. The inset of Fig.~\ref{figA2} shows, as expected, $\chi$ increases with increasing $T$. However, it is not clear from the data whether  $\xi$ approaches infinity for any finite $T=T^*$ or not. At least it is clear that for $T\leq 100$, the critical $\gamma_c$ (assuming BKT transition exists) is certainly $<0.1$. We also investigate the dependence of $T$ on the steady state entanglement fluctuation in Appendix~\ref {fluc_scaling_gamma}.}

\subsection{Dependency of steady state entanglement entropy on asymmetry parameter $\epsilon$: \label{e0}}
\begin{figure}[!ht]
    \centering
\includegraphics[width=0.48\textwidth]{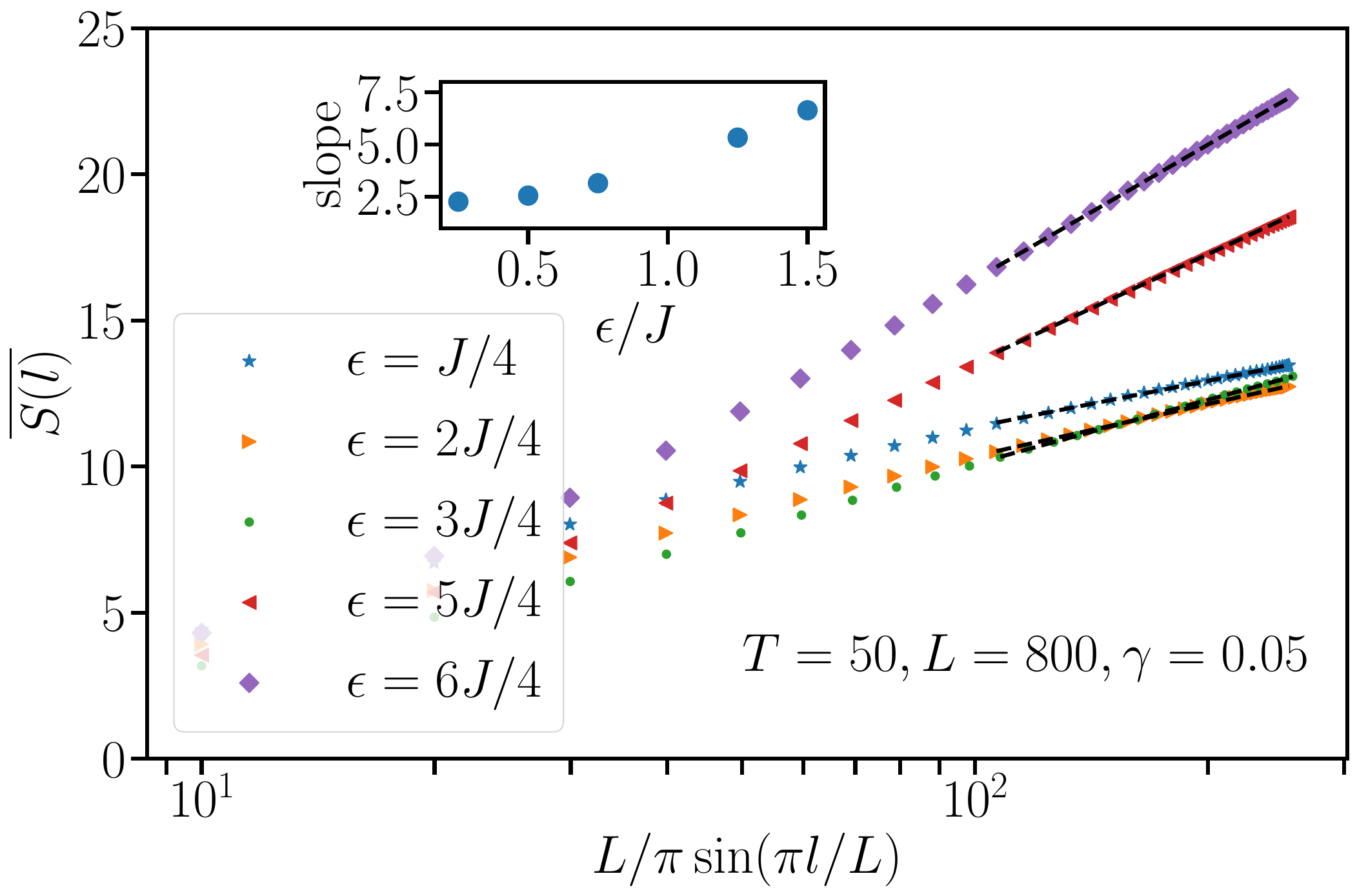}
\caption{ Comparison of entanglement entropy growth for different $\epsilon$ values with system size $L=800$ for $\gamma=0.05, T=50$. The inset shows that the slope of the entanglement growth is increasing with increasing $\epsilon$ for $T=50$.  }  
\label{entanglement_diff_epsilon}
\end{figure}

\textcolor{black}{Here we investigate the role of $\epsilon$, the asymmetry parameter in the driving amplitude in measurement-induced transition. 
In the Fig.~\ref{entanglement_diff_epsilon}, we see that the entanglement entropy slope increases with increasing $\epsilon$ for a fixed $T=50$. This also indirectly implies that if the BKT transition persists in the thermodynamic limit, then $\gamma_c$ will also increase with $\epsilon$. Similar results were obtained previously for $T=5$ as well. Next, we study the consequences of symmetric drive. }
 \begin{figure}
    \centering
    \includegraphics[width=0.48\textwidth]{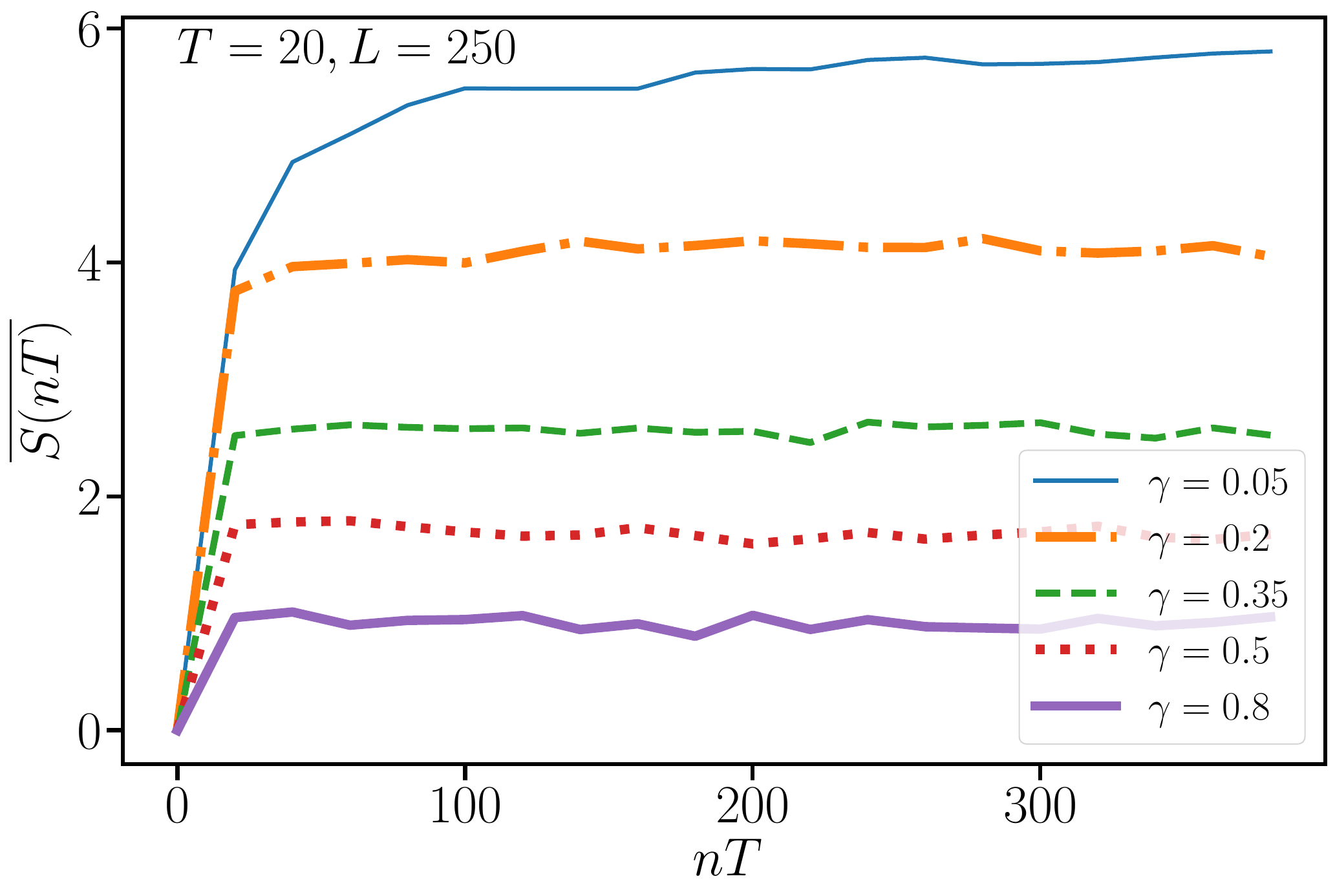}\\
    \includegraphics[width=0.48\textwidth]{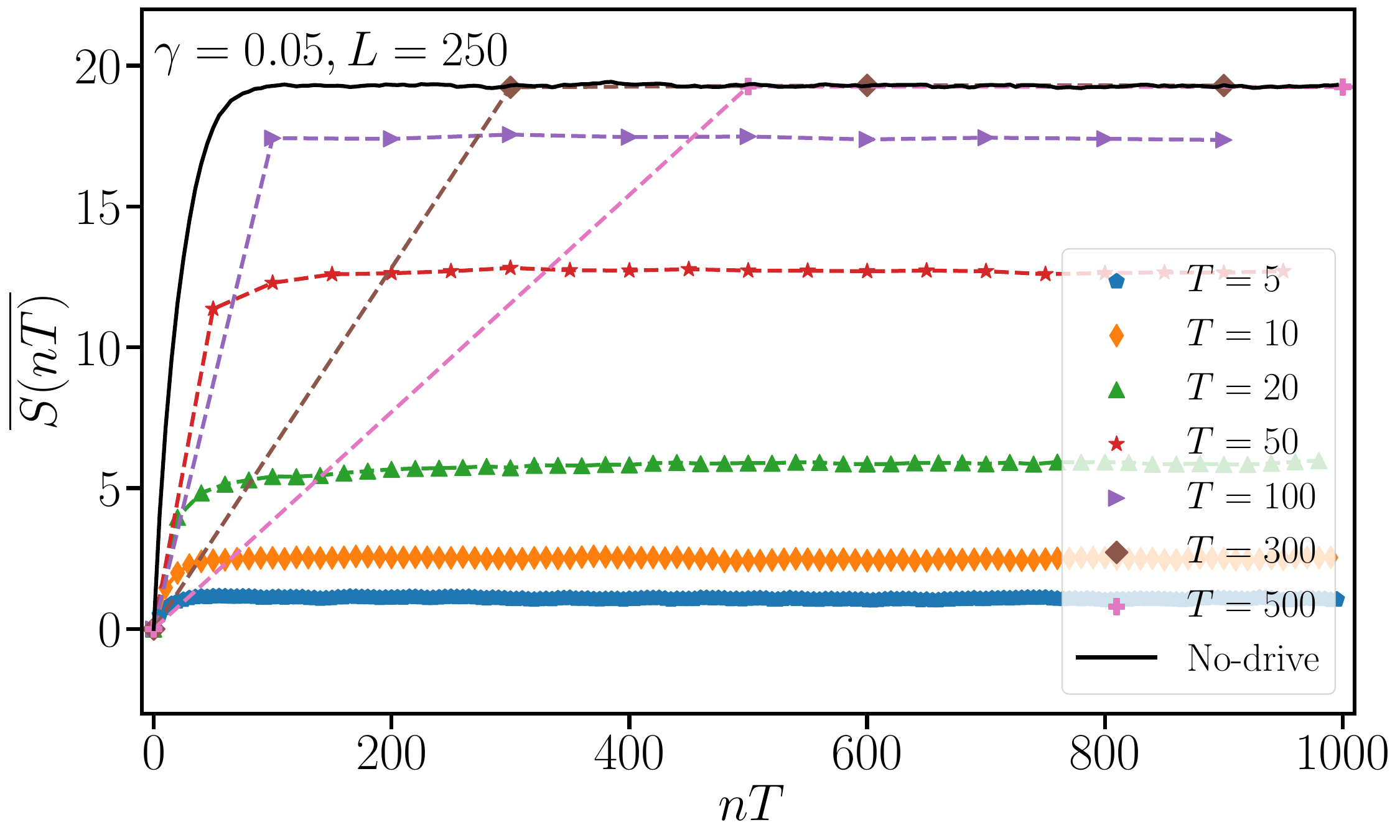}
    
    \caption{Stroboscopic growth of trajectory averaged entanglement entropy for different $T$ and $\gamma$. Upper panel: for $T=20$ and varying $\gamma$, Lower panel: for $\gamma=0.05$ and varying $T$. Plots are for system size $L=250$ and $\epsilon=0$. }
    \label{fig2}
\end{figure}
\paragraph*{Square pulse drive with zero mean: $\epsilon=0$ limit:} Here, we focus on the $\epsilon=0$ limit. First, the time evolution of the entanglement entropy, averaged over many different trajectories, is studied starting from the initial Neel state for various values of the time period $T$ and measurement strengths $\gamma$. We find that $\overline{S}$ increases initially with time and then saturates to a steady-state value. From Fig.~\ref{fig2} upper panel, we see, for a fixed driving period $T=20$, if we increase the measurement strength $\gamma$, the steady state entanglement saturation value is decreased; similar to our previous results, the entanglement spread gets suppressed because of increasing measurement strength. On the other hand, it is shown in the lower panel of  Fig.~\ref{fig2} that, for a fixed measurement strength $\gamma=0.05$, the entanglement saturation value increases with the increase of the driving time-period $T$. This suggests, in general, that the low driving frequency promotes the depinning of the particles, and the entanglement entropy saturation values approach the no-drive result with increasing $T$. We also noticed there is a critical driving period $T_c$, such that for $T>T_c$, the results are almost indistinguishable from those of the no-drive scenario. However, this $T_c$ typically increases with $L$. The fact that $T>T_c$ resembles a no-drive situation has also been observed for our $2\times 2$ Toy model previously. 
 
 Next, we immediately ask the following question: How does the steady-state entanglement entropy change with the subsystem size for different $T$? The existence of a $T_c$ for each $L$ might be a tempting scenario to conjecture that in the very large $T$ limit, $\epsilon=0$ results should resemble no-drive results and should display an entanglement transition for finite-size systems. 
However, one needs to take the large $T$ and large $L$ limits carefully.  The order of these two limits doesn't commute. If one takes a large $T$ limit first, then for any finite $L$ (no matter how large it is), $T_c$ has been achieved, and hence, one will observe no-drive results. On the other hand, the correct limit is to take the large $L$ limit first, then increase the $T$. 
\textcolor{black}{In that case, for a given $T$ and a given $\gamma$ up to a system size $L=L_1$,  the finite $T$ data will be same as the no-drive results.  However, for $L>L_1$, the result will deviate from the no-drive result since their $T_c$ has not been reached. Clearly if $T_c$ increases with $L$, taking
$L\to \infty$ first indicates that there cannot be any $T_c$ at all, so the results will never match the no-drive case.}
\begin{figure}
    \centering
    \includegraphics[width=0.48\textwidth]{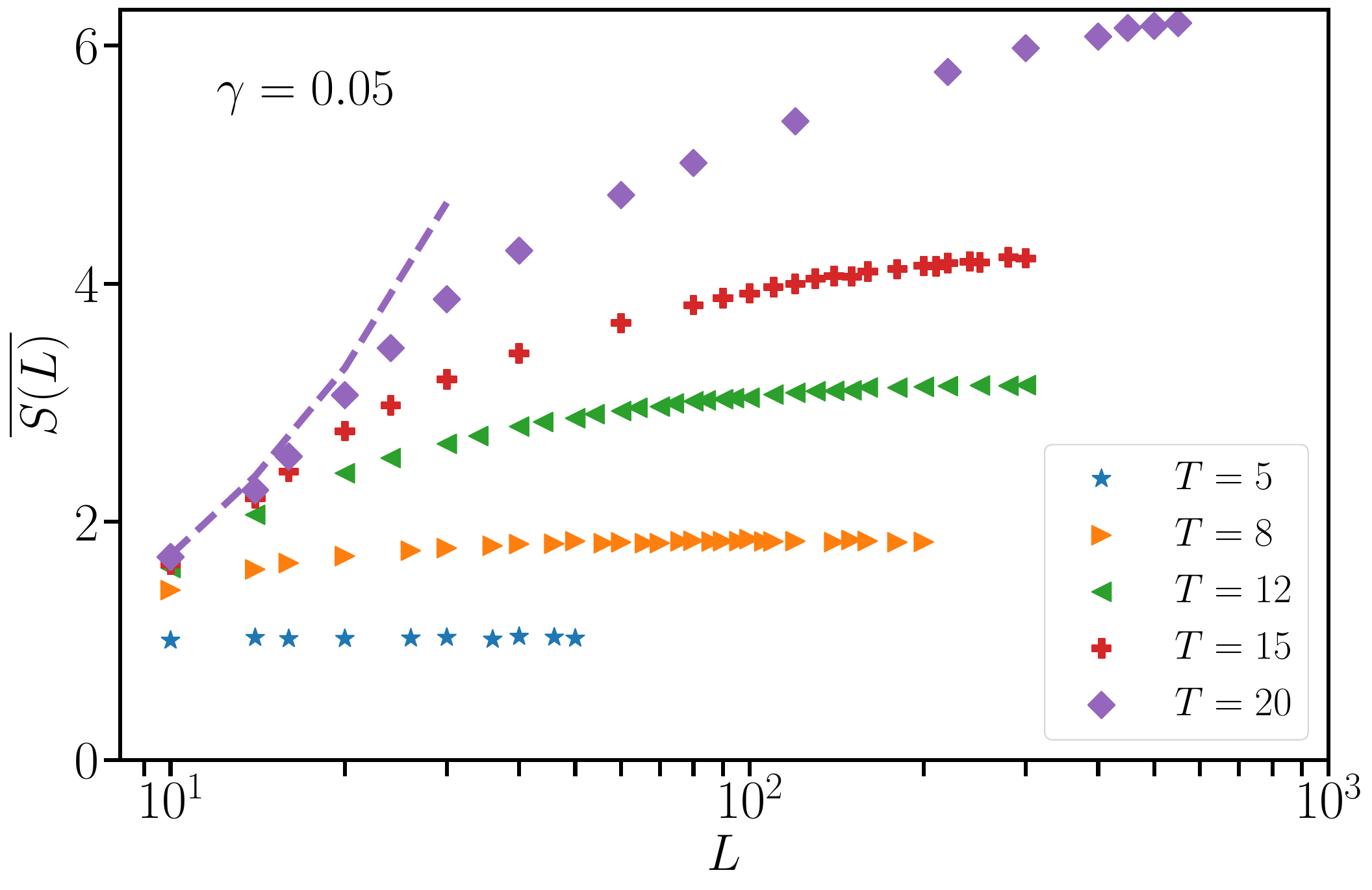}
     \includegraphics[width=0.48\textwidth]{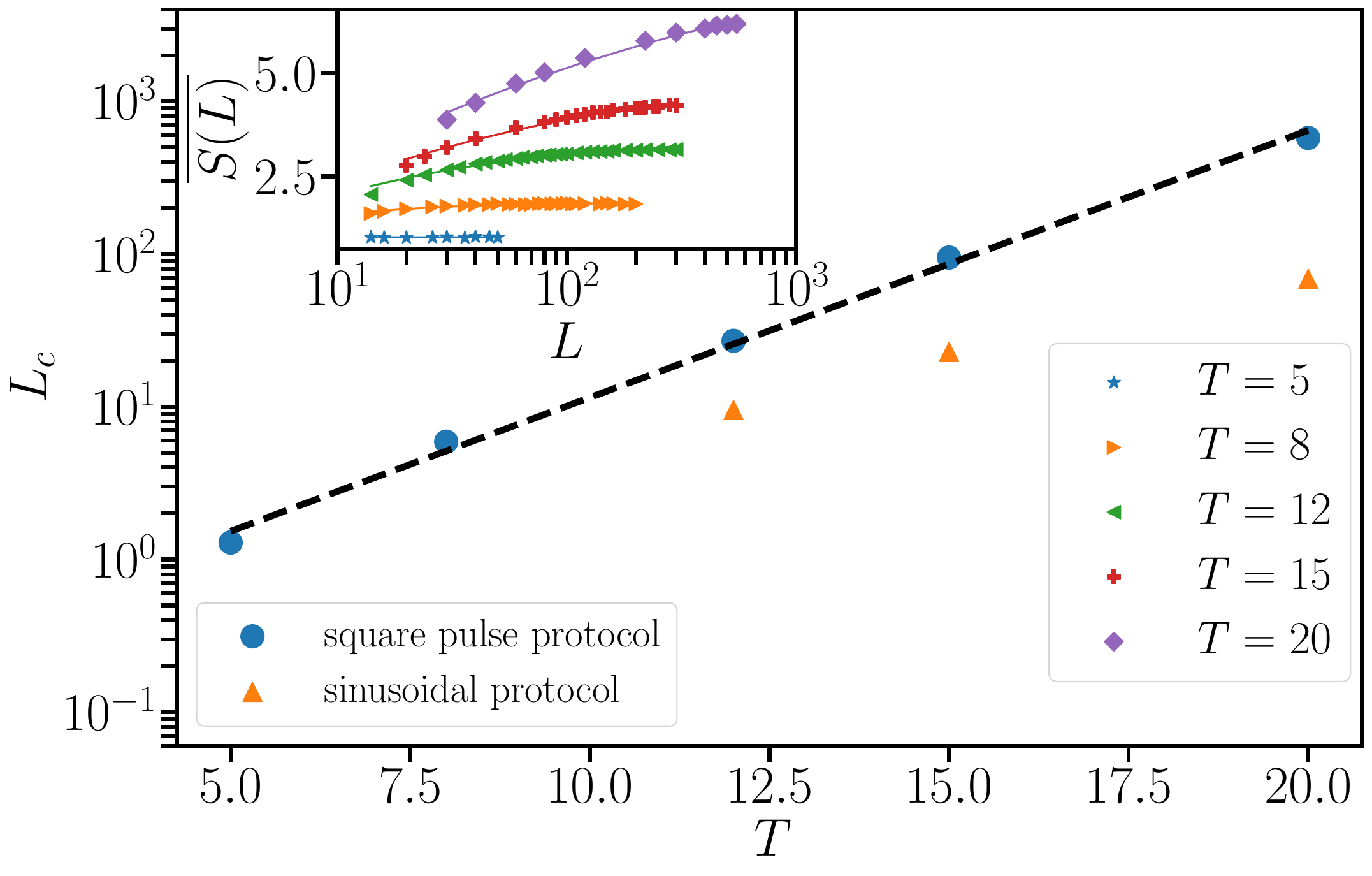}
    \caption{Upper panel: Scaling of the steady state entanglement entropy with system size $L$ for different $T$, but fixed $\gamma=0.05$ and $\epsilon=0$. Initially, it shows logarithmic growth with $L$ and then tends to saturate. 
    Dashed-line corresponds to no-drive $J=1/2$ results for $\gamma=0.05$.
    Lower panel: plot of $L_c$ vs. $T$, $L_c$ is estimated using Eq.~\eqref{fitting}, circles are for square pulse protocol, and triangles are for sinusoidal protocol. The inset shows the best fit of the entanglement entropy assuming the functional form given by Eq.~\eqref{fitting} for the square pulse protocol.}
    \label{fig5}
\end{figure}

We precisely see that 
 in Fig.~\ref{fig5} upper panel. 
  We find for $\gamma=0.05$ and  $T=5$, and $8$, the steady state entanglement entropy is almost constant for all $L$, which suggests an area law scaling with the system size. On the other hand, as we increase $T$, we see an initial logarithmic scaling of entanglement entropy with system size till $L<L_1$, followed by bending trend,  finally after a critical $L=L_c$, entanglement starts showing area law behavior. The dashed line corresponds to the no-drive results for $J=1/2$ and $\gamma=0.05$. Roughly, till $L\simeq 20$, this no-drive results resembles $T=20$ data, beyond $L=20$, the no-drive results keep increasing with $L$, while the $T=20$ data starts bending (see appendix.~\ref{appen_A} for $T=100$ results).   
  To estimate the $L_c$, we fit the steady entanglement entropy vs $L$ data with a two parameters ($S_{sat}$ and $L_c$) fitting function $f(L)$, i.e., 
 \begin{equation}
     f(L)= S_{sat} \tanh \bigg(\frac{\ln L}{\ln L_c}\bigg),
     \label{fitting}
 \end{equation}
in the inset of Fig.~\ref{fig5} lower panel. 
 In the limit $L>>L_c$,  $f(L)\to S_{sat}$, which is a constant suggesting area-law phase, on the other hand, for $L<<L_c$,  $f(L) \propto \log L$. We estimate $L_c$ from the best fit and plot $L_c$ vs $T$ in Fig.~\ref{fig5} lower panel, which suggests an exponential growth of $L_c$ with the driving period $T$.  As an exercise, we estimate the fitting parameter $L_c$ for $T=100$ and $\gamma=0.05$ from the exponential fit, which gives $L_c\simeq10^{16}$. 
 Hence, it is almost impossible to reach the accessible system size regime to see area law scaling for $T=100$ (see Fig.~\ref{fig3} in Appendix~\ref{appen_A} for details). 
 
\begin{figure}[!ht]
\centering
\includegraphics[width=0.48\textwidth]{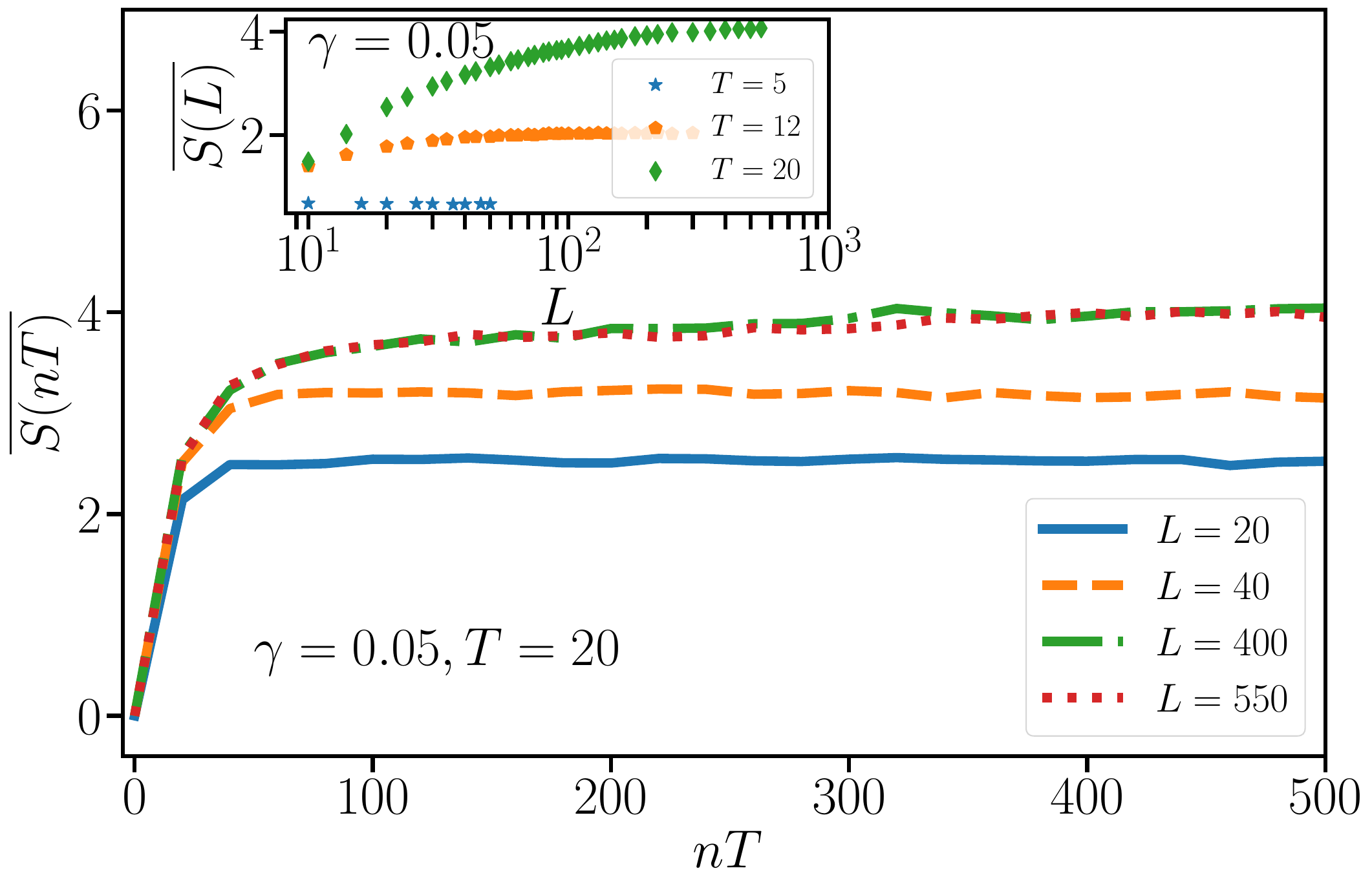}
    \caption{Stroboscopical growth of trajectory averaged entanglement entropy for different $L$ for the sinusoidal driving protocol for $T=20$ and $\gamma=0.05$(main panel). The inset shows the system size scaling of the steady state trajectory-averaged entanglement entropy for the different driving periods $T$. The data presented here are for $\gamma=0.05$.  }
    \label{fig7}
\end{figure}

 \textit{Sinusoidal drive:} 
To become more certain about our findings for the $\epsilon=0$ limit, we also study the effect of sinusoidal driving, where $J(t)=J\sin(2\pi t/T)$. We find that even for the sinusoidal drive protocol, the area-law entanglement phase sustains in the large $L$ limit. Growth of trajectory averaged entanglement entropy for $T=20$ and $\gamma=0.05$ are shown in Fig.~\ref{fig7}  for different $L$. We find that $L=400$ and $550$ data are almost indistinguishable, though for small $L$, the saturation values are different. The inset of Fig.~\ref{fig7}
shows the variation of the steady state entanglement entropy with $L$ for different $T$, and it clearly shows the evidence of the area law in the large $L$ limit, as already observed for the square pulse case.  We also estimate the $L_c$ using the fitting function $f(L)$ (see Eq.~\eqref{fitting}) and compare it with square pulse results in Fig.~\ref{fig5} lower panel. We find that $L_c$ is smaller for the sinusoidal drive compared to the square pulse. 

\section{Conclusions\label{conclusions}}
In this work, we investigate the effect of a periodic drive in a continuously monitored free-fermionic lattice model. We find that in the high-frequency limit $T<1$, the results are almost indistinguishable to the no-drive results for the effective hopping strength $J_{eff}=\epsilon/2$ at least for our accessible system size. With the growing belief of the absence of measurement-induced phase transition in the no-drive scenario~\cite{arealaw_1_starchl2024generalized, arealaw_2_muller2025monitored}, it is tempting to conclude that in the high-frequency limit, maybe we don't have any measurement-induced transition in the thermodynamic limit. No-drive results are presented in the Appendix~\ref{nodrive_central_charge} for our protocol. \textcolor{black}{It is essential to note that the numerical data are not conclusive enough to confirm the absence of measurement-induced transitions in non-interacting lattice systems like ours, probably because of the exponentially diverging correlation length predicted from NLSM.} On the other hand, we find that increasing $T$ tends to promote entanglement growth, and for reasonably large $T$ (in our manuscript, we primarily focus on $T=5$), the data suggest that the measurement-induced transition belonging to the BKT universality class may prevail even in the thermodynamic limit. However, in the absence of any concrete theoretical prediction, it is almost impossible to rule out the possibility that for extremely large system sizes, the steady-state entanglement data will keep showing area law scaling, and whatever we are observing in numerics is just a finite-size crossover between the logarithmic-area law phase. Even in this case, it is safe to say that with increasing $T$, the finite size crossover length scale will certainly increase, and for any reasonably large $T$, such system sizes will be almost inaccessible for any practical purposes~\cite{koh2023measurement}. 
We also investigate a situation where we take a reasonably large $\gamma$ for which the $T = 5$ data of the effective local central charge shows bending in the large subsystem size regime, and then we increase $T$. As expected, we could reduce the bending rate with increasing $T$, but were unable to achieve a scenario where the effective local central charge remains constant. Hence, the analysis does not allow us to conclude whether there is a finite critical driving period $T^*$ above which the bending completely stops. The absence of bending in the effective central charge data implies a critical phase. Hence, we also can’t rule out the possibility that beyond a specific critical measurement strength, the system will always remain in the area law phase for any finite driving period.

\textcolor{black}{In this context, we also want to point out the many-body localized (MBL) systems. For a typical Heisenberg model, it was initially believed that the MBL transition was at disorder strength $W\simeq 3.5$~\cite{PhysRevB.91.081103}. Then, there was a debate regarding the survival of the MBL phase in the thermodynamic limit~\cite{PhysRevE.102.062144}, and now it is believed that at least for $W<16$, there is no "true" MBL transition in the thermodynamic limit~\cite{mbl_d1,mbl_d2}, however, the most of the cold-atom experiments only can simulate the Heisenberg type model up to $L=20$ spins, and hence, for all practical purposes, the experiments only can probe that thermodynamically pre-thermal MBL phase, where dynamics is extremely slow~\cite{abanin2019colloquium}. Hence, even if it ultimately turns out that the measurement-induced transition in driven non-interacting systems is not a true thermodynamic transition, but rather a finite-size crossover, at least it is clear that the crossover length scale increases with increasing driving period, hence, even such crossover will be viewed as effectively measurement induced transition in almost all experimental platforms~\cite{exp_mipt1,exp_mipt2,exp_mipt3}.}

Moreover, we have investigated a drive protocol in which the hopping amplitudes are varied completely symmetrically around zero in the form of square-pulse and sinusoidal drives and found that such a drive can't cause a phase transition independent of any frequency regime; it always promotes an area-law entanglement phase in the thermodynamic limit. This result is a bit surprising, especially given that disorder-induced similar-driven systems tend to favor de-localization in the low-frequency regime~\cite{PhysRevB.103.184309}. Our results suggest
several interesting directions for future work. The role of disorders, interactions in the monitored Floquet system, and their relationship to Floquet-MBL~\cite{jakub.23} remain a challenging problem.  It would be interesting if one could extend the analysis presented in Ref.~\cite{mirlin.2023,arealaw_1_starchl2024generalized} to driven systems like ours in future studies.

\section{Acknowledgements}
R.M. acknowledges the DST-Inspire fellowship by the Department of Science and Technology, Government of India,
SERB start-up grant (SRG/2021/002152). The authors also thank 
M Buchhold for fruitful discussions over private communication on the RG scheme. The authors also thank Xiangyu Cao for the fruitful discussion.

\appendix
\section{$T=100$ result for square pulse protocol}\label{appen_A}
\begin{figure}[!ht]
\centering
 
    \includegraphics[width=0.48\textwidth]{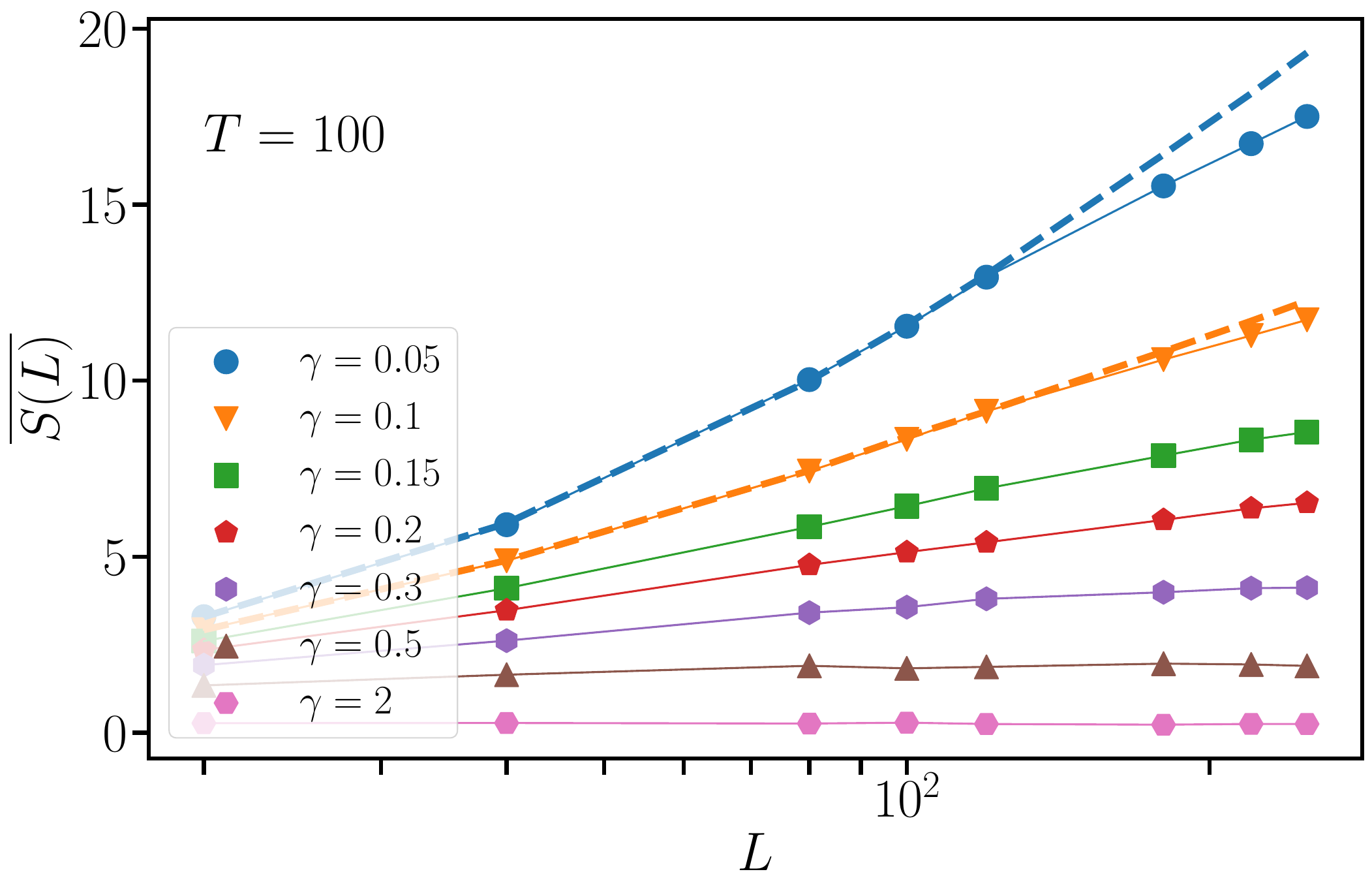}
    \caption{Variation of steady state entanglement entropy with system size for different $\gamma$ values for $T=100$ for square pulse driving protocol with $\epsilon=0$. The dashed lines show the no-drive ($J=1/2$) results.}
    \label{fig3}
\end{figure}
Here, we discuss the $T=100$ results. In Fig.~\ref{fig3}, the dashed lines correspond to no-drive results for $J=1/2$. It shows that almost up to $L\simeq 100$, the no-drive data is the same as the $T=100$ data. Afterward, $T=100$ data starts bending, while no-drive data keeps on increasing. In the main text, we show that in the case of square pulse protocol and a given driving time period $T$, for $L>>L_c$, the steady state entanglement entropy does not change with $L$. 
The value of $L_c$ was estimated using fitting function ~\eqref{fitting}. It turns out that $L_c$ increases exponentially with $T$, the fitted function predict the $L_c$ to be $~10^{16}$ for $T=100$ and $\gamma=0.05$. 
 This suggests that due to our limitations of the accessible system size,  we will be unable to identify the area-law phase for $\gamma=0.05$ and $T=100$. As expected, Fig.~\ref{fig3} shows no trace of area law for $\gamma=0.05$. The data can mislead us to predict the usual BKT type critical to area law entanglement phase transition as we increase $\gamma$.

\section{Cost-function minimization technique}\label{cost}
\begin{figure}[!ht]
\centering
\includegraphics[width=0.48\textwidth]{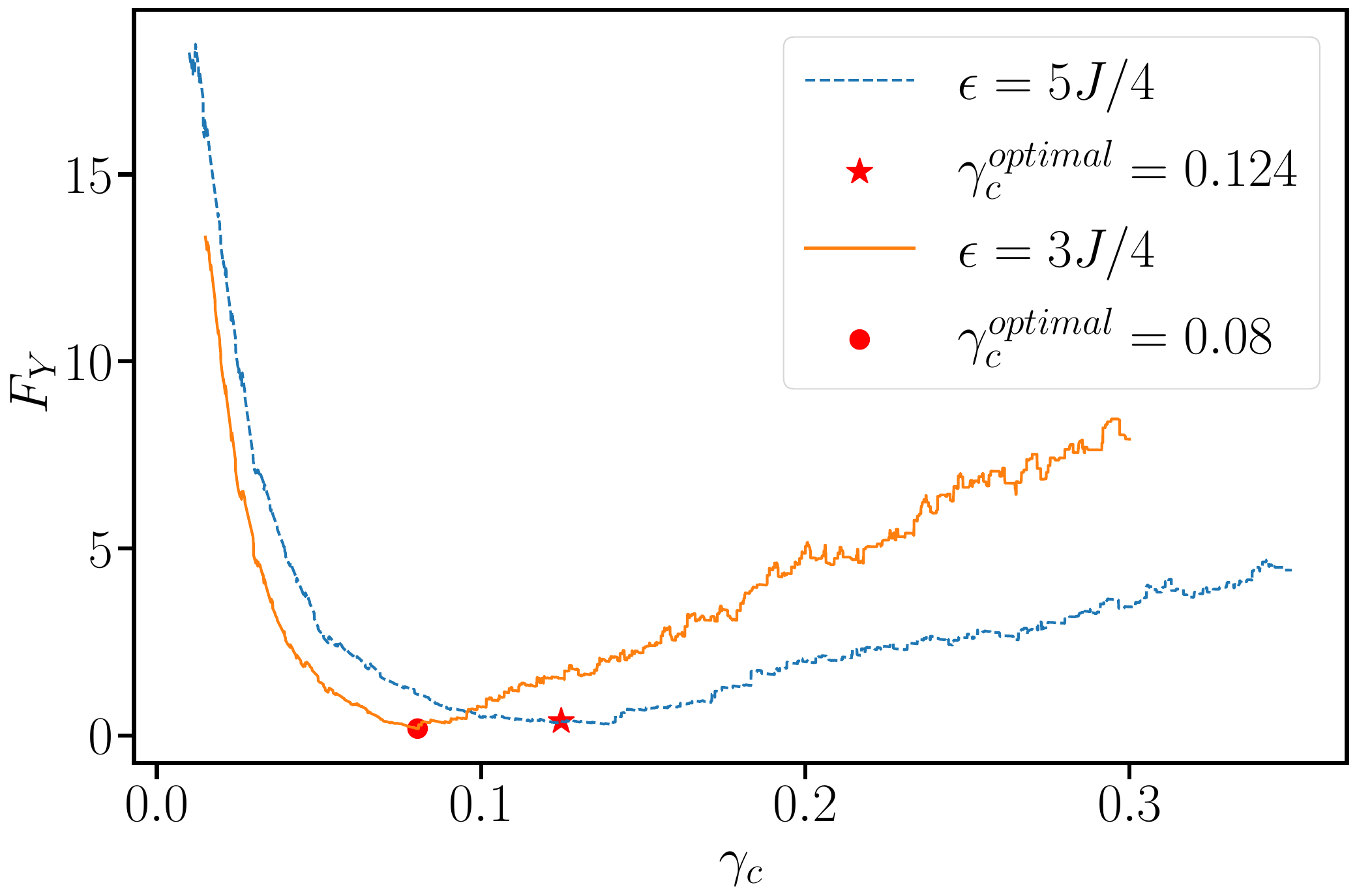}
     \caption{Cost function $F_Y$ vs. $\gamma_c$ plot for different $\epsilon$ values. The star represents the optimal  $\gamma_c$ for $\epsilon=5J/4$, and the circle represents the optimal $\gamma_c$ value for $\epsilon=3J/4$. }
  \label{cost_fn}  
\end{figure}
\begin{figure}[!ht]
\centering
\includegraphics[width=0.48\textwidth]{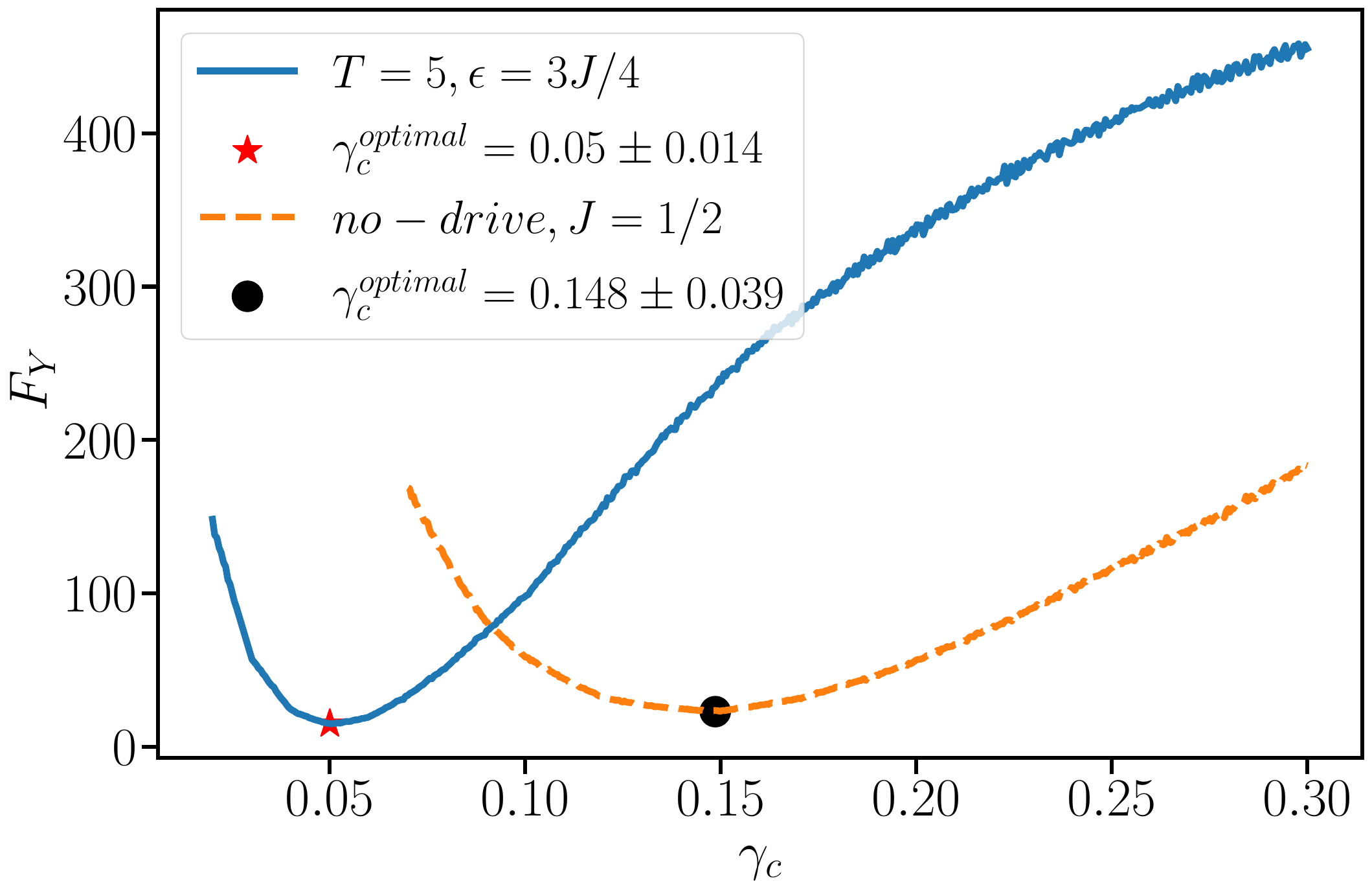}
     \caption{Cost function $F_Y$ vs. $\gamma_c$ plot following the scaling Eq.~ \ref{new_entangle_data_collapse} for  $\epsilon =3J/4, T=5$ and no drive $J=1/2$. The red star represents the optimal  $\gamma_c$ for $\epsilon=3J/4$, and the black circle represents the optimal $\gamma_c$ value for $J=1/2$ no drive. }
     \label{cost_new}
\end{figure}
In order to see the BKT transition, we had to perform a finite-size scaling analysis.
The job is to find out the $\gamma_c$ for which one can get the best possible data-collapse as observed in Fig.~\ref{fig_new_b} (lower panel). 
We take $X_i = (\gamma-\gamma_c )(\ln L)^2$ for all the values of $\gamma$ and $L$, and $Y_i = \overline{S(L, \gamma)}-\overline{S(L, \gamma_c)}$ sorted according to the increasing values of $X_i$. The values of $\overline{S(L, \gamma)}$, which are not in the list, have been calculated using linear \textcolor{black}{interpolation}\cite{PhysRevB.102.054302,interpolation_PhysRevX.9.031009}. Now, one can define the cost function as~\cite{PhysRevB.102.064207},
\begin{align*}
    C_Y=\frac{\sum_{i=1}^P |Y_{i+1}-Y_{i}|}{\max\{Y_i\}-\min\{Y_i\}}-1,
\end{align*}
where $P$ is the total number of unique $X_i$ present in the list. For a perfect data-collapse, $C_Y=0$. We perform a minimization of the function $C_Y$ with respect to the free parameter $\gamma_c$ and let for $\tilde{\gamma}_c$ the cost function is minimum, i.e., $C_Y(\tilde{\gamma}_c)=(C_Y)_{\text{min}}$ (see Fig.~\ref{cost_fn}). Usually, such minima in $C_Y$ may not be that robust, by changing the input data set (adding or removing some data), minima can shift a bit. 
To incorporate that effect, we identify a set of $\{\gamma_c\}$ 
for which $C_Y \leq 2C_Y(\tilde{\gamma}_c)$. Then identify $\gamma^{(1)}_c=\text{min}\{\gamma_c\}$ (minimum from that set) and $\gamma^{(2)}_c=\text{max}\{\gamma_c\}$ (maximum from that set). 
We identify the optimal $\gamma_c$ as $\gamma_c=(\gamma^{(1)}_c+\gamma^{(2)}_c)/2$ (similar technique was used in Ref.~\cite{areejit_prb} to compute the error-bars).

\textcolor{black}{To obtain the $\gamma_c$ using the scaling Eq.~\ref{new_entangle_data_collapse} we have used $X_i = (\gamma-\gamma_c )(\ln ({\frac{L}{\pi}\sin(\frac{\pi l}{L}))})^2$ and $Y_i = \overline{S(l, \gamma)}-\overline{S(l, \gamma_c)}$. The cost function values and optimal $\gamma_c$ values are presented in Fig.~\ref{cost_new}.}

\section{F-test}\label{f-test}
We perform the $F$-test to quantitatively determine the statistical likelihood of our steady state entanglement data presenting a logarithmic scaling or volume law for small $\gamma$. 
The $F$-test is a likelihood-ratio test that assesses the goodness of fit of two competing statistical models~\cite{mood1950introduction, fazio.prbl.24}.
For each set of $\overline{S(L)}$ , we produce two curves $\tilde{S}_{L}$ and $\tilde{S}_{\ln(L)}$, which are linear best-fits curves of $\overline{S(L)}$ in $L$ and $\ln(L)$ respectively.
These best-fit lines are used to compute the $F$-test statistic,
\begin{equation}
    F_T = \frac{\tilde{E}_{L}}{\tilde{E}_{\ln(L)}}
,\end{equation}
where $\tilde{E}_{L}$ ($\tilde{E}_{\ln(L)}$) is the sum of the squared errors between $\overline{S(L)}$ and $\tilde{S}_{L}$ ($\tilde{S}_{\ln(L)}$).
The way $F_T$ is constructed, a lower $F_T$ will favor volume law, and a higher $F_T$ will favor logarithmic scaling. We exercise this test for our $\overline{S(L)}$ vs $L$ data and found the $F_T>>1$ for $T=5$, $20$, $100$, $150$, $200$ and  for $\gamma=0.05$, and $\epsilon=3J/4$. It confirms that the scaling is most likely logarithmic(for finite size). 

\section{Effective central charge in the absence of drive}\label{nodrive_central_charge}

\begin{figure}[!ht]
    \centering
    \includegraphics[width=0.48\textwidth]{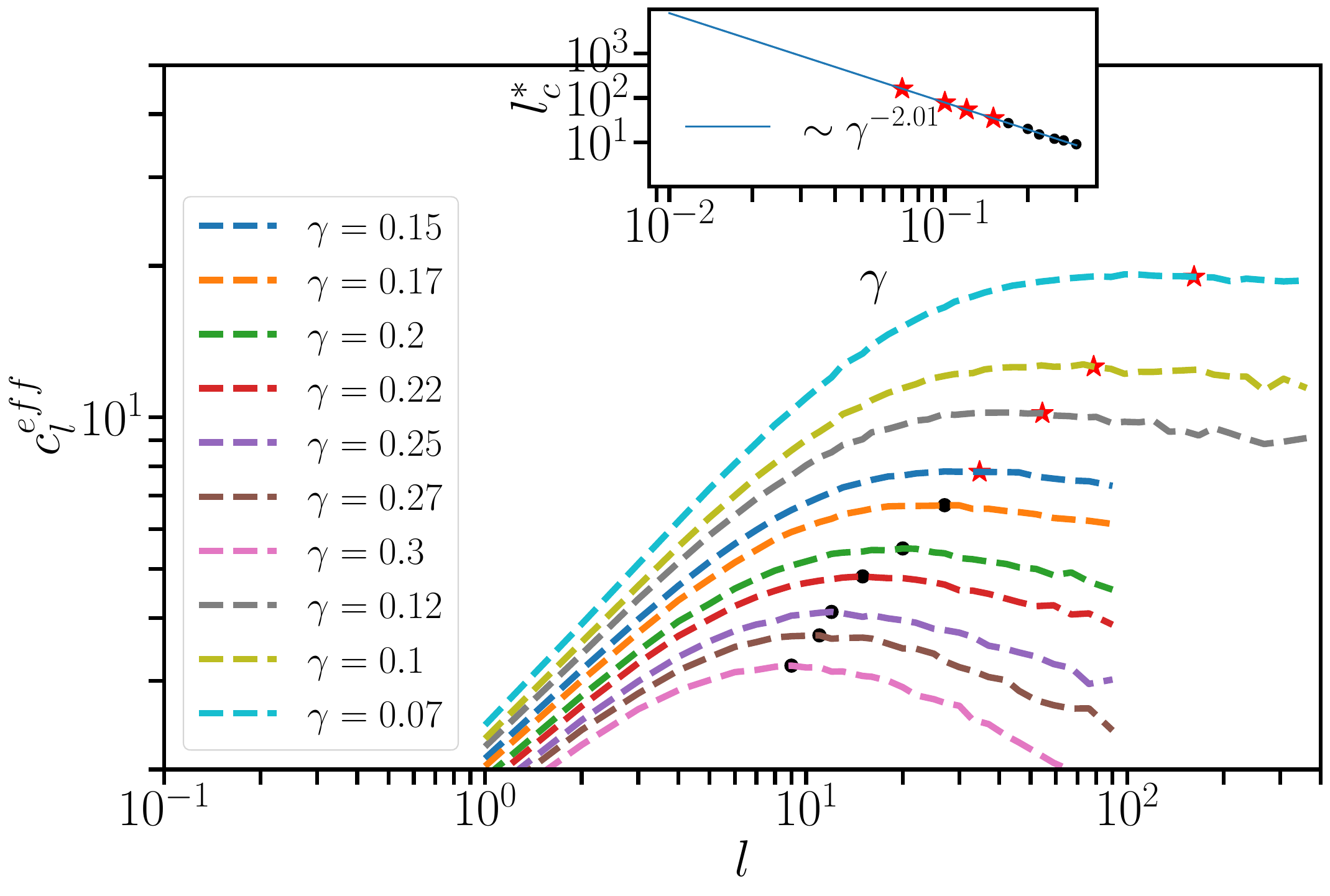}
    \caption{Effective central charge $c^{eff}_l$ vs. $l$ plots for different $\gamma$ values for nodrive, $J=1/2$. Black circular points represent the maximum of the central charge and red star points represent the position of $l^*_c$ for which $c^{eff}_l$ is the maximum estimated from the fitting. Inset shows the best fit, which gives $l^*_c \sim \gamma^{-2.01}$. The maximum system size is taken as $L=1004$.  }
    \label{nodrive_effective}
\end{figure}

\textcolor{black}{In this section, we present the results for the effective central charge in the absence of driving with $J=1/2$. 
Though these are the no-drive results, for the accessible system sizes in our study, the high-frequency limit ($T < 1$) results should resemble them. This is because the entanglement entropy data in that limit is almost indistinguishable from the no-drive data with $J_{eff} = \epsilon/2$. 
Using the same cost function minimization technique, we get the transition point $\gamma_c=0.148\pm0.039$. We calculate the $l^*_c$ for which $c^{eff}$ is maximum for some $\gamma> \gamma_c$, and best fit gives $l^*_c \sim \gamma^{-2.01}$, from the fitting we calculate the $l^*_c$ values and try to match with the actual simulated data with system size $1004$ for $\gamma \le \gamma_c$ as we did in the main text for $T=5$. In Fig.~\ref{nodrive_effective}, we observe evidence of bending for $\ \gamma=0.12$ and $ 0.1$, but for $\gamma=0.07$, the bending is not very clearly evident. Note that in Ref.~\cite{starchl2024generalized}, a similar analysis was done for a different measurement protocol.}

\section{QR decomposition and normalization in each trotter step}
\label{QR decomposition}

\textcolor{black}{
The problem discussed in the paper
$U$ evolves as,
\begin{equation}
    U(t+dt)=e^Me^{-iH(t)dt}U(t),\nonumber
\end{equation}
where $H(t)=J(t)\sum_{i=1}^L (\hat{c}_i^{\dag}\hat{c}_{i+1} + h.c.)
$ is the single-particle Hamiltonian. Under this non-hermitian evolution, the isometry
$U(t+dt)^\dagger U(t+dt)= \mathcal{I}_{N\cross N}$ does not satisfy. To recover this isometry property, we do $QR$ decomposition as,
$U(t+dt)=QR$. Where $ Q$ is an $L \cross N$ matrix with orthonormal columns, satisfying
  \begin{equation}
      Q^\dagger Q = \mathcal{I}_{N \times N}.\nonumber
  \end{equation}
and $ R $ is an $ N \times N $ upper triangular matrix.
Since \( Q \) retains the orthonormality condition, we replace \( U(t+dt) \) with \( Q \), ensuring:
\begin{equation}
    U(t+dt) \rightarrow Q, \quad \text{where } Q^\dagger Q = \mathcal{I}_{N \times N}.\nonumber
\end{equation}
Thus, the QR decomposition projects \( U(t+dt) \) back onto the manifold of isometric matrices, maintaining the normalization of the quantum state. 
\begin{figure}[!ht]
    \includegraphics[width=0.48\textwidth]{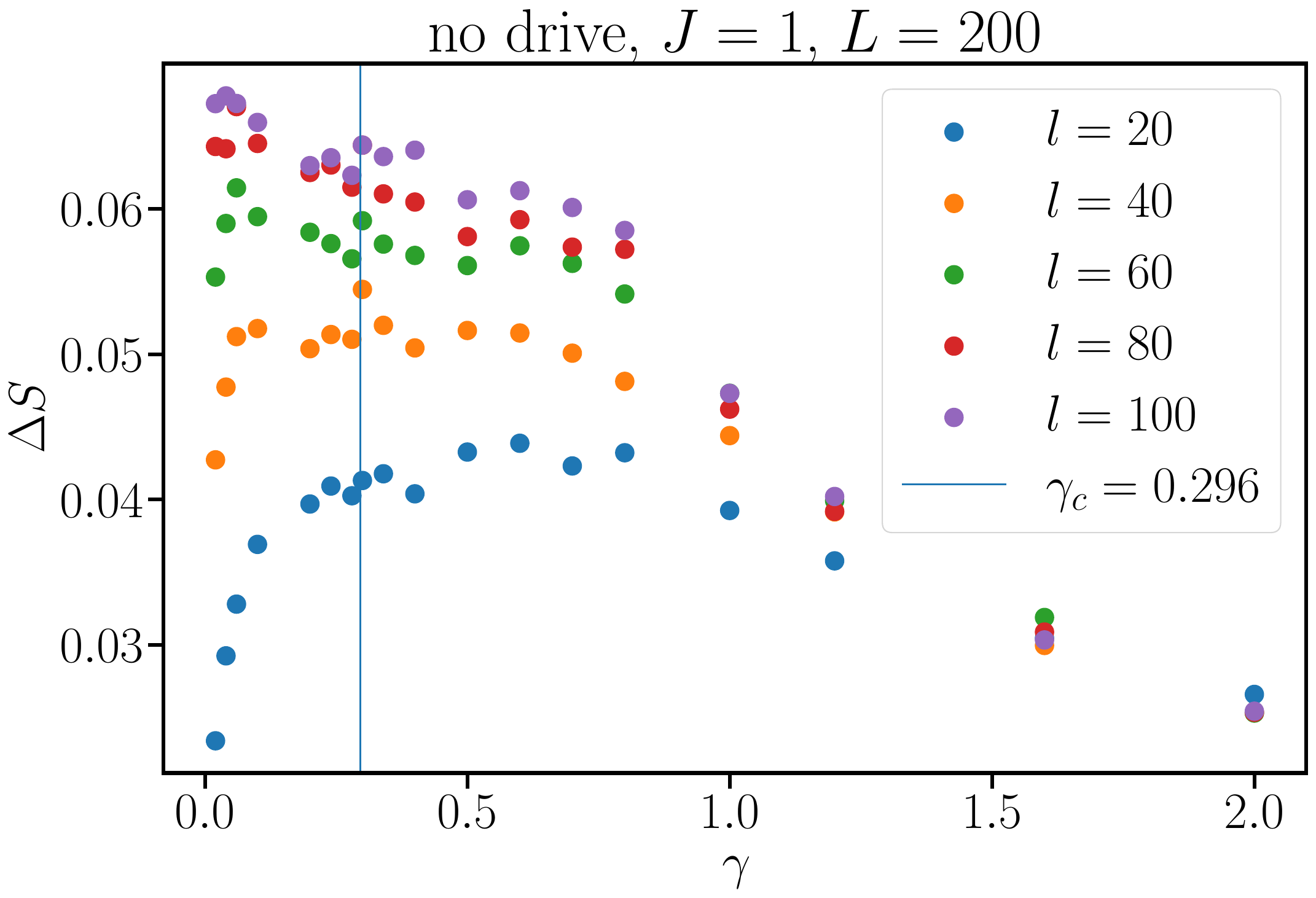}
\includegraphics[width=0.48\textwidth]{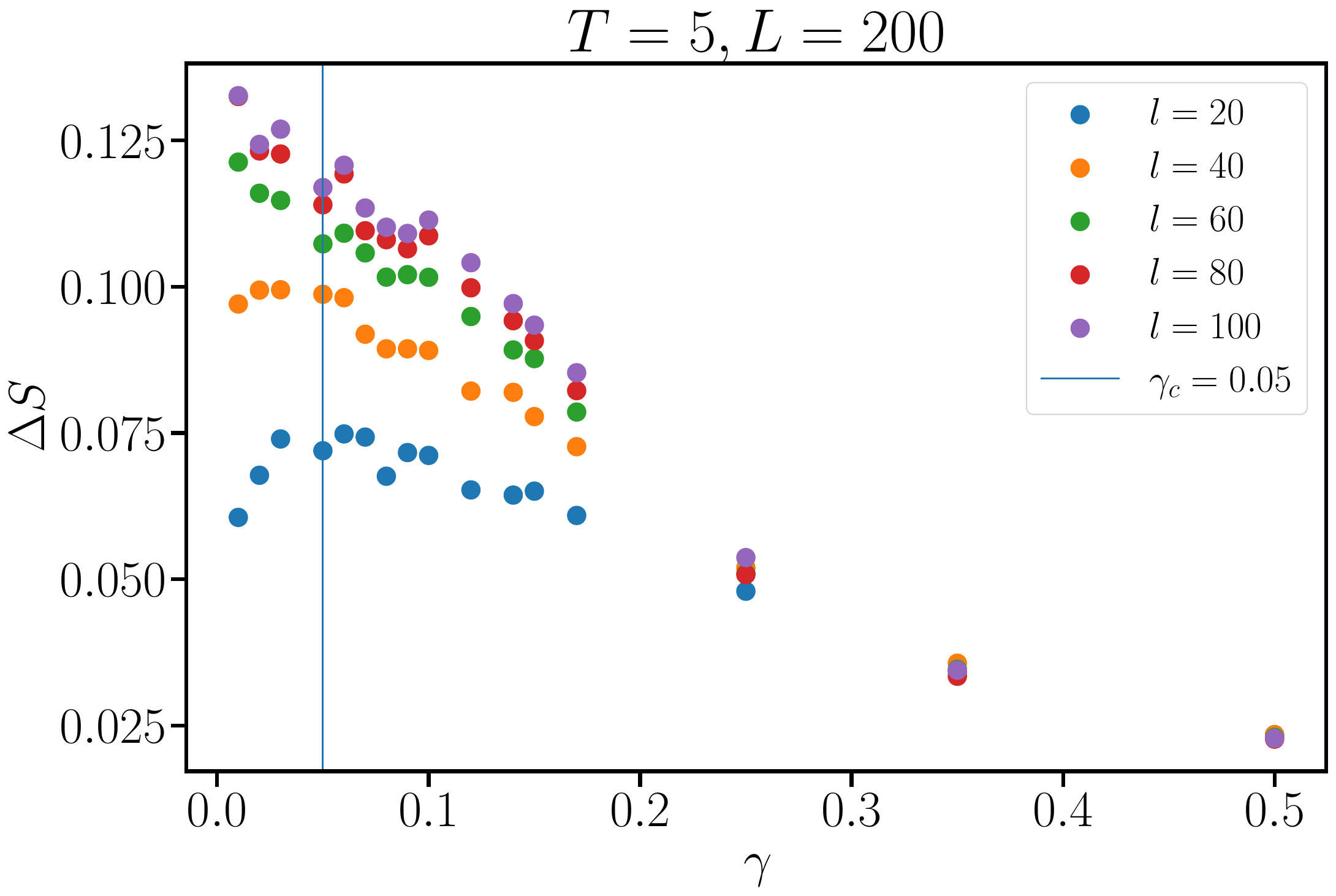}
\caption{Fluctuation vs. $\gamma$ for different subsystem size $l$ for no drive(Upper panel) and $T=5$ case(lower panel).}
\label{fluc_gamma_fig}
\end{figure}
Since the columns of \( Q \) remain orthonormal, it ensures that the quantum state remains normalized.}

\section{Fluctuation of steady state entanglement entropy}\label{fluc_scaling_gamma}
\begin{figure}
    \centering
    \includegraphics[width=0.48\textwidth, height=0.25\textwidth]{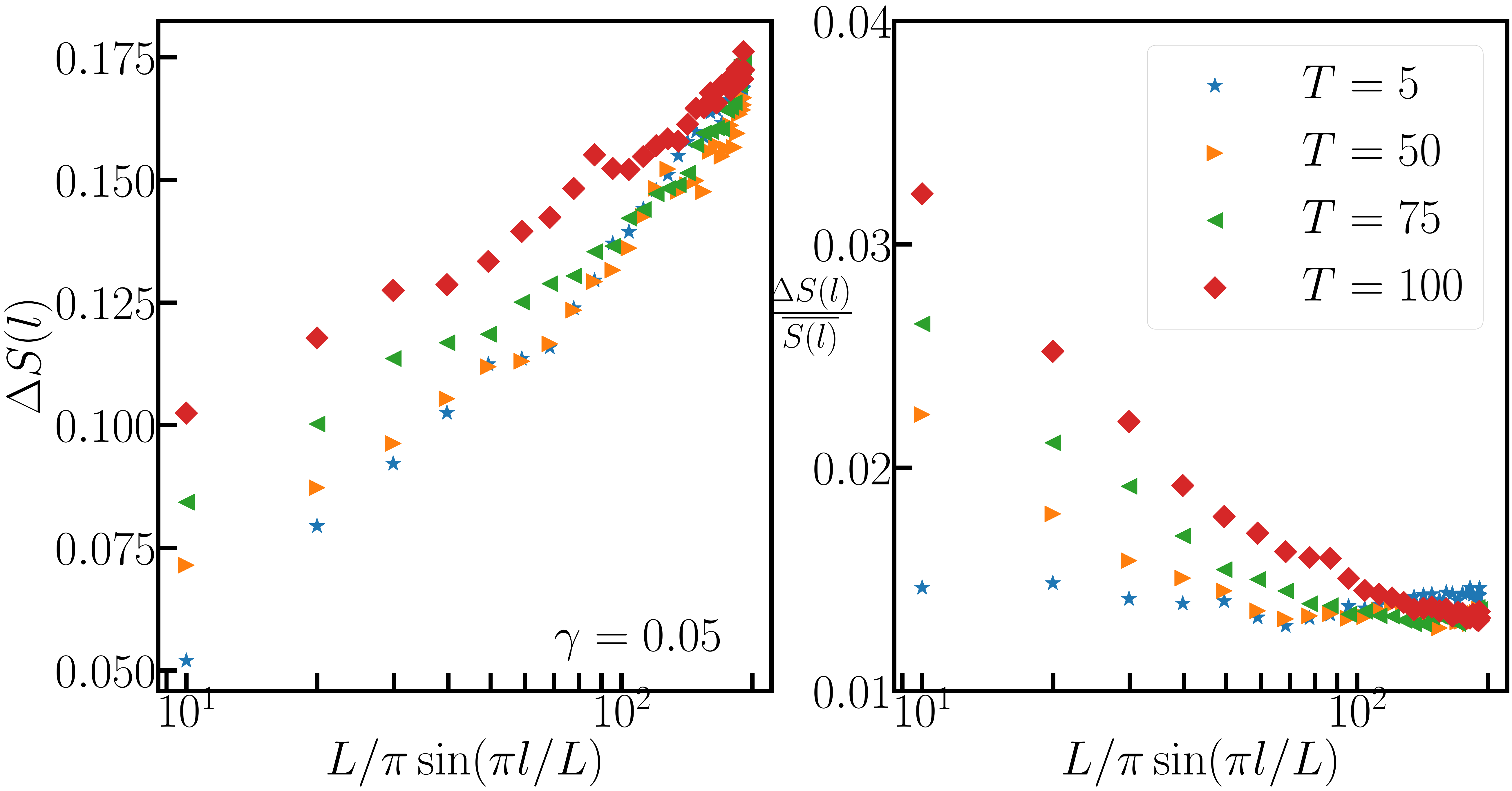}
    \includegraphics[width=0.48\textwidth, height=0.25\textwidth]{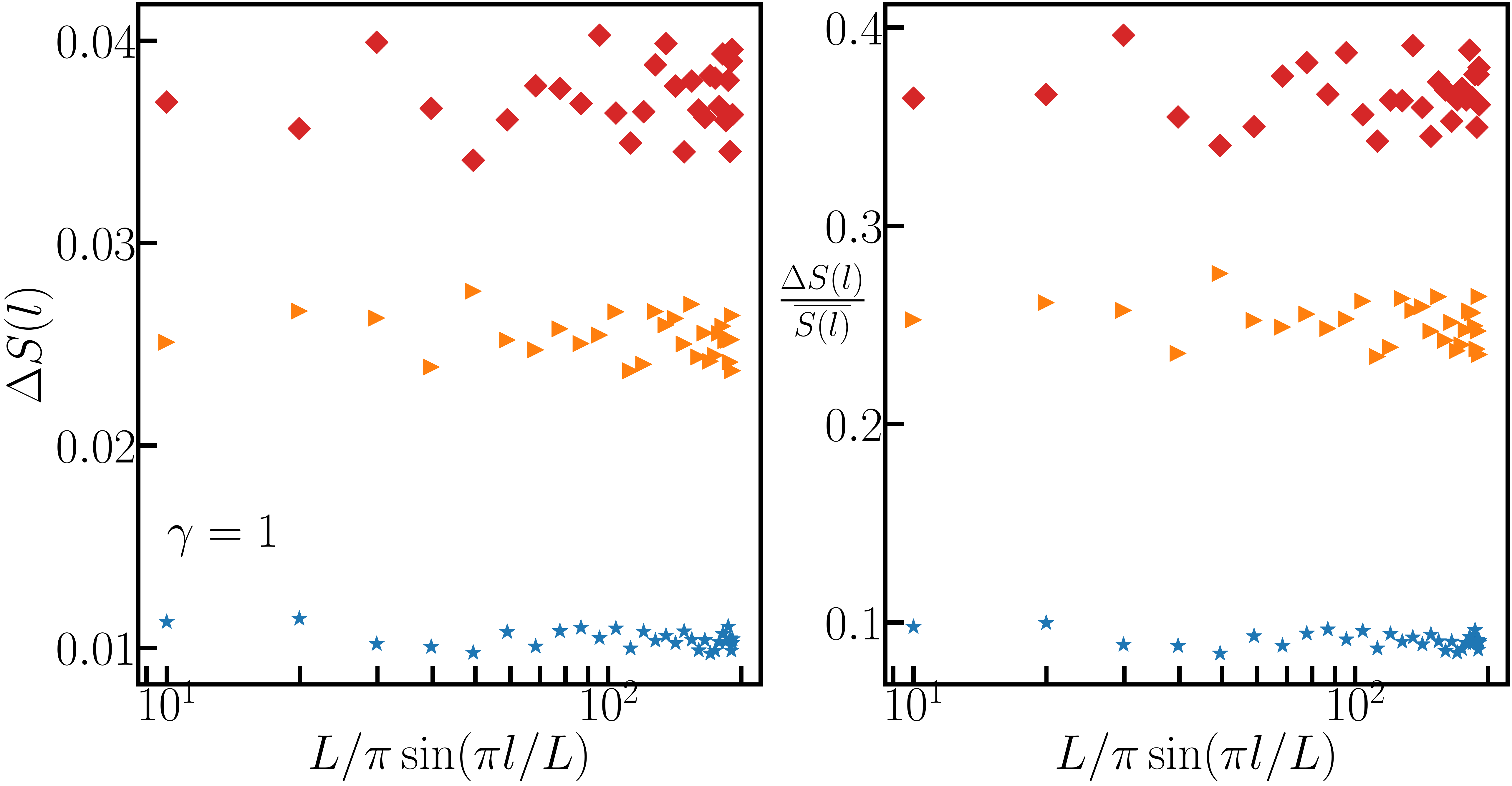}
    \caption{Upper left panel: Scaling of standard deviation/fluctuation of long time averaged entanglement entropy over various trajectories $\Delta S$ with subsystem size $l$ for different driving periods $T$ for $\gamma=0.05$. Upper right panel: fluctuation-mean ratio vs cord length for different driving periods. System size $L=600$ for $\gamma=0.05$. The lower panel presents the result for $\gamma=1$.}
    \label{entanglement_fluc}
\end{figure}
\textcolor{black}{In this section, we have calculated the standard deviation of the time-averaged entanglement entropy for various different trajectories $\Delta S=\sqrt {{\overline { S^2}}-(\overline{ S })^2}$, where the overline represents the trajectory average and $S$ is the long time averaged(taken stroboscopically) entanglement entropy value of each trajectory. . 
From Fig.~\ref{fluc_gamma_fig} we see for small $l$ (sub-system size), the fluctuation $\Delta S$ tends to show a peak near the critical $\gamma_c$; however, for reasonably large $l$, that peak seems to disappear and the $\Delta S$ seems to decrease with $\gamma$ monotonically. This signature prevails for the no-drive and even for the finite $T=5$ case.}

\textcolor{black} { More precisely, we investigate how the standard deviation\cite{kalsi2022three} of the long-time averaged entanglement entropy over various trajectories  behaves with system sizes for different $T$,
In Fig.~\ref{entanglement_fluc} upper left panel, we see that in general, fluctuation($\Delta S$) increases with $l$, and also for small system size, it increases with $T$; for large $l$, for different $T$, the difference between their fluctuations is getting smaller for $\gamma=0.05$. Also, we calculated the fluctuation and mean ratio in the upper right panel for $\gamma=0.05$, which decreases with $l$, and for a given small sub-system size, decreases with decreasing $T$. Again, in the large $l$ limit, the difference between the fluctuation-mean ratio gets smaller for different $T$s. On the other hand, the lower panels show in the area-law regime $\Delta S$ and the fluctuation-mean ratio both are not scaling with $l$, but they increase with driving period $T$ for $\gamma=1$.}

\section{Correlation function analysis for $T=5$}\label{corr func}
\begin{figure}
    \centering
    \includegraphics[width=0.48\textwidth]{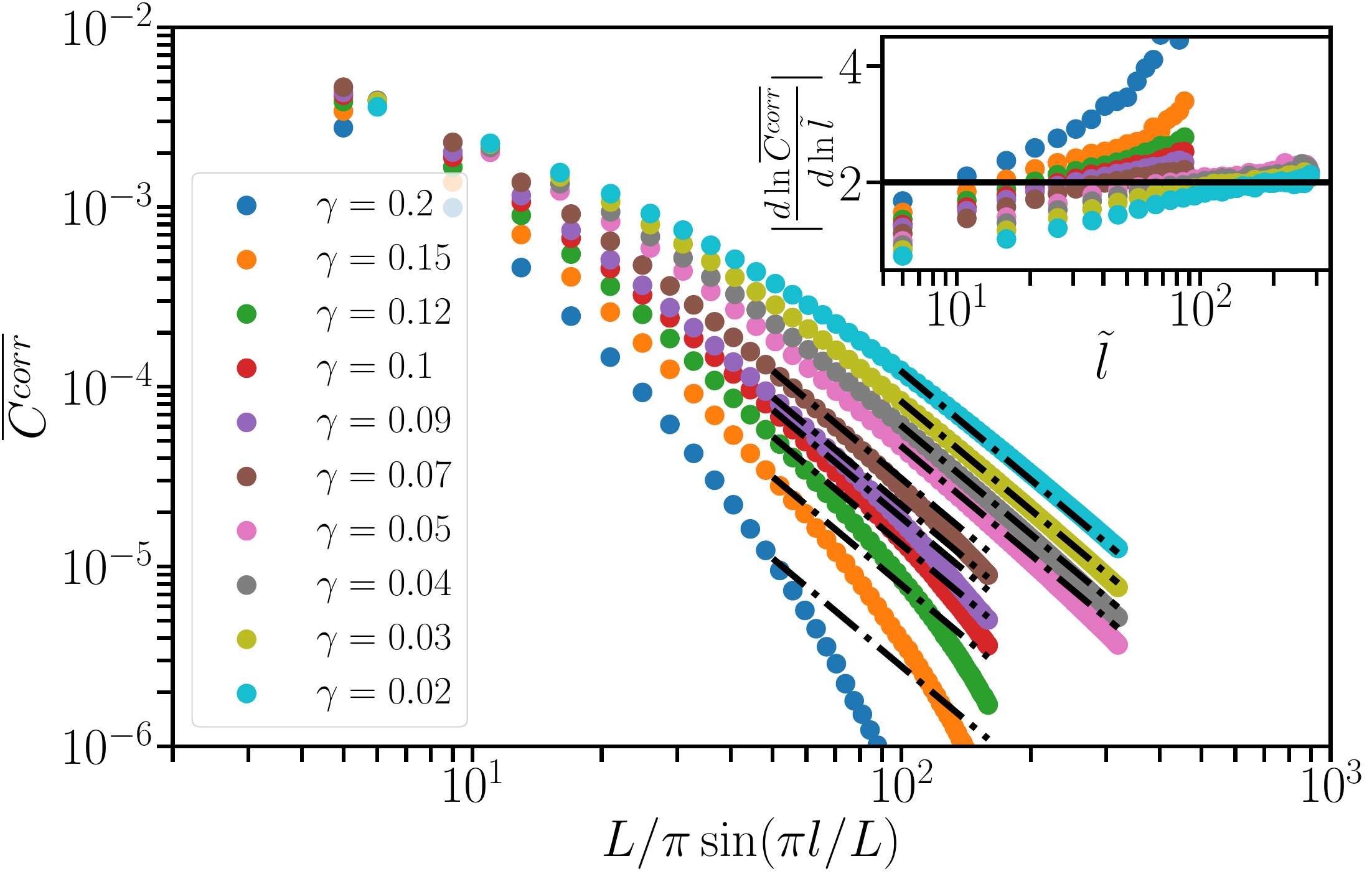}
    \caption{Equal time connected correlation function at steady state $\overline{C^{corr}}$ vs. $\tilde{l}$ plot for $T=5$. The black dashed line represents the fitting function $\overline{C^{corr}} \sim \tilde{l}^{-2}$, where $\tilde{l}=\frac{L}{\pi}\sin(\pi l/L)$. The maximum system size is taken as $L=1004$. Inset shows the variation of $|d\ln \overline{C^{corr}}/d\ln \tilde{l}|$ vs $\tilde{l}$, which seems to saturate to $2$ for $\gamma=0.03$, $0.02$ for large $\tilde{l}$.}
    \label{corr_fig}
\end{figure}

\textcolor{black}{In addition to the entanglement entropy, in the case of $T=5$, here we further calculate the equal time connected correlation function in steady state, which is defined as,
\begin{equation}
    C^{corr}(l)\equiv \overline{|D_{{l+i},i}|^2}=\overline{\langle \hat{n}_{l+i}\rangle\langle \hat{n}_i\rangle}-\overline{\langle\hat{n}_{l+i}\hat{n}_i\rangle}.
\end{equation}
Here, the overline denotes averaging over different trajectories as well as over positions ($i$) in the steady state. In the critical phase, the correlation function is expected to decay algebraically, whereas in the area-law phase, an exponential decay of the connected correlation function is anticipated~\cite{diehl_prl_2021}. For both projective and jump measurement protocols, deviations from a power-law dependence in the small $\gamma$ regime have been numerically reported in the absence of drive~\cite{mirlin.2023,starchl2024generalized}.\\
In our analysis, we compute the connected steady-state correlation function (see Fig.~\ref{corr_fig}) for $T=5$. We observe that for $\gamma < \gamma_c = 0.05 \pm 0.01$ (assuming a BKT transition exists), particularly at $\gamma = 0.03$ and $0.02$, there are no significant signs of deviation from algebraic decay within the system sizes accessible to us. This observation aligns with the entanglement entropy data. However, we cannot rule out the possibility that such deviations might emerge at much larger length scales. }

\bibliography{cite}
\end{document}